\pdfoutput=1

\documentclass{article}

\usepackage{amsmath}
\usepackage{amssymb}
\usepackage{amsfonts}
\usepackage{epsfig}

\newcounter{NN}
\newtheorem{proposition}[NN]{Proposition}
\newtheorem{theorem}[NN]{Theorem}
\newtheorem{corollary}[NN]{Corollary}
\newtheorem{lemma}[NN]{Lemma}
\newtheorem{definition}[NN]{Definition}
\newtheorem{remark}[NN]{Remark}

\def\proof{\noindent {\bf Proof:} }
\def\qed{{\hfill $\square$\\ \noindent}}

\def\:{{,}}

\def\r#1#2{{\cal R}(#1,#2)}
\def\ROOT#1{\Phi_{#1}}
\def\PROOT#1{\Phi^\prime_{#1}}
\def\abs#1{{|#1|}}

\def\C{\mathbb{C}}
\def\R{\mathbb{R}}
\def\Z{\mathbb{Z}}
\def\N{\mathbb{N}}
\def\g{{\mathfrak g}}
\def\AR{{\mathcal A}}
\def\L{{\mathcal L}}
\def\G{{\mathcal G}}
\def\B{{\mathcal B}}
\def\H{{\mathcal H}}
\def\M{{\mathcal M}}

\def\vec#1#2{\left( \begin{array}{c} #1 \\ #2 \end{array} \right)}
\def\mat#1#2#3#4{\left( \begin{array}{cc} #1 & #2 \\
                                          #3 & #4 \end{array} \right)}

\begin{document}
\bibliographystyle{plain}
\title{Global classification of two-component
approximately integrable evolution equations}
\author{Peter H.~van der Kamp}
\date{Department of Mathematics, \\
La Trobe University, Victoria, 3086, Australia\\[3mm]
Email:\ P.vanderkamp@latrobe.edu.au }

\pagestyle{plain} \maketitle

\begin{abstract}
We globally classify two-component evolution equations, with
homogeneous diagonal linear part, admitting infinitely many
approximate symmetries. Important ingredients are the symbolic
calculus of Gel'fand and Diki{\u\i}, the Skolem--Mahler--Lech
theorem, results on diophantine equations in roots of unity by F.
Beukers, and an algorithm of C.J. Smyth.
\end{abstract}

\section{Introduction} \label{intro}
A long standing open problem is the classification, up to linear
transformations, of two-component integrable equations
\begin{equation} \label{twoc}
\vec{u_t}{v_t} = \vec{a u_n + F(u,v,u_1,v_1,\ldots)}{b v_n +
G(u,v,u_1,v_1,\ldots)}
\end{equation}
where $F,G$ are purely nonlinear polynomials in variables $u_i,v_i$,
which denote the $i$-th $x$-derivatives of $u(x,t),v(x,t)$.
Among the many different approaches to recognition and classification
of integrable equations, the so called symmetry approach has proven
to be particularly successful, see for example \cite{MSS, TsWo} and references
in there.
Until recently, all results obtained were for classes of equations at fixed
(low) order $n$. This situation changed dramatically when, by
using a symbolic calculus and results from number theory, Sanders and
Wang classified scalar evolution equations with
respect to symmetries globally, that is, where the order $n$ can be
arbitrarily high \cite{MR99g:35058}. Our aim is to obtain a similar result for
the class of multi-component equations (\ref{twoc}).

In the symmetry approach the existence of infinitely many generalized
symmetries is taken as the definition of integrability. A generalized
symmetry of equation (\ref{twoc}) is a pair of differential polynomials
$S=(S_1,S_2)$ such that equation (\ref{twoc}) is also satisfied by
$\tilde{u}=u+\epsilon S_1$, $\tilde{v}=v+\epsilon S_2$ up to order
$\epsilon^2$. This leads to the notion of Lie-derivative: $\L(K)S=0$
$\Leftrightarrow$ $S$ is a symmetry of $(u_t,v_t)=K$.

The Lie algebra of pairs of differential polynomials is
a graded algebra. The linear part $(au_n,bv_n)$ has total grading
$0$, the quadratic terms have total grading $1$, and so on. Gradings
are used to divide the condition for the existence of a symmetry into
a number of simpler conditions: $\L(K)S\equiv 0$ modulo quadratic
terms, $\L(K)S\equiv 0$ modulo cubic terms, and so on. This has been
called the perturbative symmetry approach \cite{psa}. In the same
spirit the idea of an approximate symmetry was defined \cite{MNW}. If
$\L(K)S\equiv 0$ modulo cubic terms, we say that $S$ is an
approximate symmetry of degree 2. And, we call an equation approximately
integrable if it has infinitely many approximate symmetries.

We contribute to the above mentioned problem by globally classifying
equations (\ref{twoc}) that are approximately integrable  of degree
2. This is achieved by applying the techniques developed in the
special case of so called $\B$-equations, where any approximate
symmetry of degree 2 is a genuine symmetry \cite{PH02}. It extends
older results obtained by Beukers, Sanders and Wang
\cite{MR99i:35005,MR1829636}. See \cite{top} for an overview on the
application of number theory in the analysis of integrable evolution
equations and \cite{MNW2} for more recent results. The present article
is a revised and extended version of the report \cite{prep}.

As remarked in \cite{MNW} the requirement of the existence of approximate
symmetries of degree $2$ is very restrictive and highly non-trivial.
On the other hand an equation may have infinitely many approximate
symmetries of degree $2$, but fail to have any symmetries. This problem
involves conditions of higher grading and is left open.

\section{Generalized symmetries} \label{gesy}
A symmetry-group transforms one solution to an equation to another
solution of the same equation. We refer to the book of Olver
\cite{Olv} for an introduction to the subject, numerous examples and
applications.

We denote $\AR=\C[u,v,u_1,v_1,\ldots]$ and $\g=\AR\otimes\AR$. We
will endow $\g$ with the structure of a Lie algebra. For any
$K=(K_1,K_2)\in\g$ the pair $S=(S_1,S_2)\in \g$ is a {\it
generalized symmetry} of the two-component evolution equation
\begin{equation} \label{eve}
\vec{u_t}{v_t}=\vec{K_1}{K_2}
\end{equation}
if the {\it Lie derivative of $S$ with respect to $K$},
\begin{equation} \label{lib}
\L(K)S=\vec{\delta_K(S_1)-\delta_S(K_1)}{\delta_K(S_2)-\delta_S(K_2)},
\end{equation}
vanishes. Here $\delta_Q$ is the prolongation of the evolutionary
vector field with characteristic $Q$, cf. \cite[equation 5.6]{Olv},
\[
\delta_{(Q_1,Q_2)}=\sum_{k=0}^\infty D_x^k Q_1
\frac{\partial}{\partial u_k} + D_x^k Q_2 \frac{\partial}{\partial
v_k},
\]
and the {\it total derivative} $D_x$ is\footnote{In \cite[section 4.1]{top}
the total derivative was denoted $\delta_x$. This is misleading as
$D_x=\delta_{(u_1,v_1)}$. Also $\delta_Q$ is the unique $\C$-linear
derivation on $\AR$ satisfying $\delta_Q(u,v)=Q$ {\em and} $\delta_Q \circ
D_x = D_x \circ \delta_Q$.}
\[
D_x=\sum_{k=0}^\infty u_{k+1} \frac{\partial}{\partial u_k} + v_{k+1}
\frac{\partial}{\partial v_k}.
\]
The Lie derivative is a representation of $\g$. This property, with
$P,Q\in\g$,
\begin{equation} \label{rep}
\L(\L(P)Q)=\L(P)\L(Q)-\L(Q)\L(P)
\end{equation}
corresponds to the Jacobi identity for the Lie bracket $[P,Q]=\L(P)Q$
which is clearly bilinear and antisymmetric, cf. \cite[Proposition
5.15]{Olv}. Another way of expressing (\ref{rep}) is saying that $\g$
is a $\g$-module. Another $\g$-module is given by $\AR$, the
representation being $\L(K)F=\delta_K(F)$ with $K\in\g,F\in\AR$.

The word `generalized' stresses the fact that the order of a symmetry can
be bigger than one. Generally symmetries come in hierarchies
with periodic gaps between their orders. For example, the Korteweg--De Vries
equation $u_t=u_3+uu_1$ possesses odd order symmetries only. Concurrently,
the KDV equation has approximately symmetries at any order.

\section{Grading} \label{grad}
Denote $\sigma_u=(u,0)$ and $\sigma_v=(0,v)$. If $P$ in some
$\g$-module is an eigenvector of $\L(\sigma_{u})$ (or of
$\L(\sigma_{v})$), the corresponding eigenvalue is called the $u$-
(or $v$-) {\it grading} of $P$. If $P$ has $u$-grading $i$ and $v$-grading
$j$ we say that $i+j$ is the {\it total grading} of $P$.
One verifies that $\g$ can be written as the direct sum
\[
\g=\bigoplus_{t\geq 0} \g^t,\quad \g^t=\bigoplus_{-1 \leq i\leq t+1}\g^{i,t-i},
\]
where elements of $\g^{i,j}$ have $u$-grading $i$ and $v$-grading $j$.
For example, $(u_1v_2,v_3v_4)\in\g^{0,1}$ has total grading 1.
Similarly we have
\[
\AR=\bigoplus_{t\geq 0}\AR^t,\quad \AR^t=\bigoplus_{0\leq i\leq t}\AR^{i,t-i}.
\]

The crucial property of a graded Lie algebra is that the $u$-, (or
$v$-, or total) grading of $\L(P)Q$ is the sum of the $u$-, (or $v$-, or total)
gradings of $P$ and $Q$. This follows directly from equation (\ref{rep}).
Gradings are used to divide the condition for the existence of a symmetry
into a number of simpler conditions.

We study evolution equations of the form
\begin{eqnarray}
\vec{u_t}{v_t}&=&K^0 + K^1 + \cdots \notag \\
&=& K^{0,0} + K^{-1,2}+K^{0,1}+K^{1,0}+K^{2,-1} + \cdots, \label{eveq}
\end{eqnarray}
with $K^{0,0}=(a u_n,b v_n)$ and symmetries of similar form
$S=S^{0} + S^{1} + \cdots$ with $S^{0} = S^{0,0} = (c u_m, d v_m)$.\footnote{We remark that only if $a=b$ then $S$ may
also contain terms $S^{\pm 1,\mp 1}$. In this paper we implicitly assume this
does not happen.} Here the dots may contain terms with total
grading $>1$. Certainly we have $\L(K^{0,0})S^{0,0}=0$.
The symmetry conditions with total grading 1 are
\begin{eqnarray}
&\L(K^{-1,2})S^{0,0}+\L(K^{0,0})S^{-1,2}&=0,\nonumber\\
&\L(K^{0,0})S^{0,1}+\L(K^{0,1})S^{0,0}&=0,\nonumber\\
&\L(K^{0,0})S^{1,0}+\L(K^{1,0})S^{0,0}&=0,\label{eqs}\\
&\L(K^{0,0})S^{2,-1}+\L(K^{2,-1})S^{0,0}&=0.\nonumber
\end{eqnarray}
An implicit function theorem which, under certain conditions,
guarantees that $S$ is a symmetry of $K$ if the first few symmetry
conditions of total grading $0,1,\ldots$ are fulfilled, was given by
Sanders and Wang \cite{MR99g:35058,SW01a}. In this paper we restrict
ourselves to solving equations (\ref{eqs}). Thus we classify
equations that admit infinitely many approximate symmetries of
degree 2, which is a necessary condition for integrability. In the
sequel we omit the adjective `of degree 2'.

\section{The Gel'fand--Diki{\u\i} transformation} \label{syme}
Comparing the Leibniz rule and Newton's binomial formula,
\[
(uv)_n=\sum_{i=0}^n \binom{n}{i} u_iv_{n-i},\qquad
(x+y)^n=\sum_{i=0}^n \binom{n}{i} x^iy^{n-i},
\]
we see that differentiating a product is quite similar to taking the
power of a sum. On the right hand side the index, counting the number
of derivatives, gets interchanged with the power, while on the left
hand side differentiation becomes multiplying with the sum of
symbols. Of course, with expressions containing both indices and
powers, one has to be more careful. The Gel'fand--Diki{\u\i}
transformation \cite{MR58:22746} provides a one to one correspondence
between $\AR^{i,j}$ and the space $\C^{i,j}$: polynomials in
$\C[x_1,\ldots,x_i,y_1,\ldots,y_j]$ that are symmetric in both the
$x$ and the $y$ symbols. One may deduce the general rule from
\[
u_1 u_2 v_3 \trianglerighteq \frac{x_1^1 x_2^2 + x_2^1 x_1^2}{2!}
\frac{y_1^3}{1!} = \widehat{u_1 u_2 v_3},
\]
or consult one of the papers \cite{top,psa,MNW2}. All usual
operations from differential algebra translate naturally. In
particular,\footnote{As a correction to \cite[Section 4.3]{top}, when
$(f,g)\in\g^{i,j}$ then $f\in\AR^{i+1,j}$ and $g\in\AR^{i,j+1}$. One
should think of $(f,g)$ as representing the vector field
$f\partial_u+g\partial_v$.}
\[
\L(K^{0,0})S^{i,j}\trianglerighteq
\mat{\G^{i,j}_{1;n}[a\:b]}{0}{0}{\G^{i,j}_{2;n}[a\:b]}\widehat{S^{i,j}},
\]
where the so called $\G$-functions are given by
\begin{eqnarray*}
\G^{i,j}_{1;n}[a\:b](x,y)&=&a (x_1^n+\cdots+x_{i+1}^n) + b(y_1^n+\cdots+y_j^n) \\
&&-a(x_1+\cdots+ x_{i+1} + y_1+\cdots+ y_j)^n,
\end{eqnarray*}
and
\begin{equation} \label{gequ}
\G^{i,j}_{2;n}[a\:b](x,y)=\G^{j,i}_{1;n}[b\:a](y,x).
\end{equation}
Symbolically we can solve the symmetry conditions of total grading 1, equations (\ref{eqs}), as follows.
We may write the components of the quadratic parts of $S$ as, with $k=1,2$,
\begin{equation} \label{rhs}
\widehat{S^{i,j}_k}=\frac{\G^{i,j}_{k;m}[c\:d]}{\G^{i,j}_{k;n}[a\:b]}
\widehat{K^{i,j}_k}.
\end{equation}
Equation (\ref{eveq}) has an approximate symmetry at order $m$ with linear coefficients $c,d$ iff
for all $i+j=1$ and $k=1,2$ the right hand side of equation (\ref{rhs}) is either polynomial or
undefined ($0/0$).

\section{Nonlinear injectivity} \label{ni}
In our classification we distinguish between equations whose approximate symmetries necessarily have
non-vanishing linear part and equations that allow purely nonlinear approximate symmetries.
\begin{definition}
Let $K^0$ have total grading 0. We call $K^0$ {\em nonlinear injective} if $\L(K^0)S=0$
implies that $S$ has total grading 0. And, we call an equation nonlinear injective if its
linear part is nonlinear injective.
\end{definition}
With $K^0=(au_n,bv_n)$, the $k$-th component of $\L(K^{0})S^{i,j}$, with
non-zero $S^{i,j}$, vanishes iff $\G^{i,j}_{k,n}[a\:b]=0$. Solving the later equation with $i+j=1$
yields $ab=0$, $n\geq 0$, or $n=1$, or $(a-2b)(2a-b)=0$, $n=0$. In Table \ref{tab} we have displayed
all $K^0$ and corresponding $S^1$, such that the equation $(u_t,v_t)=K^0+K^1$,
with arbitrary $K^1\in\g^1$, has purely nonlinear approximate symmetries $S^1\in\g^1$. Note that the classification is performed up to linear transformations. In particular we may interchange $u$ and $v$.
Therefore without loss of generality we set $b=1$ and classify the values of $a$ up till inversion.
\begin{table}[h]
\begin{center}
\begin{tabular}{|c||c|c|c|c|c|c|}
\hline
&&&&&\\
$K^0$ & $(0,v)$ & $(2u,v)$ & $(au_1,v_1), a\neq1$ & $(u_1,v_1)$ & $(0,v_n)$, $n>1$ \\
&&&&&\\
\hline
&&&&&\\
$S^1$ & $\g^{1,0}$ & $\g^{-1,2}$ & $ \AR^{2,0}\otimes\AR^{0,2}$ & $\g^1$ & $ \AR^{2,0}\otimes 0$\\
&&&&&\\
\hline
\end{tabular}
\caption{\label{tab} List of $K^0$ and $S^1$ such that $\L(K^0)S^1=0$.}
\end{center}
\end{table}

For the same choices of $K^0$ and $S^1$ the linear equation $(u_t,v_t)=K^0$ has symmetries $(cu_m,dv_m)+S^1$ for
all $m\in\N$ and $c,d\in\C$. Indeed, every $\B$-equation, that is, an equation of the form (\ref{eveq}) with
$K^1\in\g^{-1,2}$, admits the zeroth order symmetry $(2u,v)$. In fact, every tuple $S\in\g$ is a symmetry of
$(u_t,v_t)=(u_1,v_1)$. Or, in other words, $(u_1,v_1)$ is a symmetry of every equation.

Only a subset of the equations $(u_t,v_t)=K^0+K^1$, with particular $K^1\in\g$,
has infinitely many symmetries with non-vanishing linear part. There is a good reason for including such equations in
the classification: their approximate symmetries may correspond to approximately integrable nonlinear injective equations.
One integrable example, equation (\ref{zjjh}), is given in section \ref{cr}. On the other hand, nonlinear injectivity is
one of the conditions in the implicit function theorem of Sanders and Wang, see section \ref{grad}.

\section{Necessary and sufficient conditions} \label{nsc}
In this section we introduce convenient notation, we give necessary and sufficient conditions for a
nonlinear injective equation to be approximately integrable, and we outline how we perform the classification.

The components of equation (\ref{eveq}) are
\begin{equation} \label{ceveq}
\vec{u_t}{v_t}=\vec{a u_n + K_1^{1,0} + K_1^{0,1} + K_1^{-1,2} + \cdots }{
b v_n + K_2^{0,1} + K_2^{1,0} + K_2^{2,-1}+ \cdots}.
\end{equation}
We denote the symbolic representation of the 6-tuple $K_1^{1,0}$,
$K_1^{0,1}$, $K_1^{-1,2}$, $K_2^{0,1}$, $K_2^{1,0}$, $K_2^{2,-1}$ by
$\widehat{K}^1$. And similarly we write $S_1^{1,0}$, $\ldots$,
$S_2^{2,-1} \trianglerighteq \widehat{S}^1$ and
$\G_n=\G_{1;n}^{1,0}, \ldots, \G_{2;n}^{2,-1}$. A $6$-tuple $H$ is
called {\em proper} if it consists of polynomials with the right symmetry
properties, that is, if
$H\in\C^{2,0}\otimes\C^{1,1}\otimes\C^{0,2}\otimes\C^{0,2}\otimes\C^{1,1}\otimes\C^{2,0}$.
Thus, $\widehat{K}^1$, $\widehat{S}^1$, and $\G_n[a\:b]$ are proper
tuples. We will also consider $s$-tuples, with $s<6$. It should be
clear from the context in which space a proper $s$-tuple lives. We
say that an $s$-tuple $H=H_{[1]},\ldots, H_{[s]}$ divides an $s$-tuple
$P=P_{[1]},\ldots,P_{[s]}$ if $H_{[i]} \mid P_{[i]}$ for all $1\leq i\leq s$ and we
write $P/H=P_{[1]}/H_{[1]},\ldots,P_{[s]}/H_{[s]}$. We are now able to state the
following: Equation (\ref{ceveq}) is nonlinear injective and has an
approximate symmetry of order $m$ with linear coefficients $c,d$ iff
the 6-tuple $\widehat{S}^1=\G_m[c\:d] \widehat{K}^1 / \G_n[a\:b]$ is
proper.

Let $H$, $\G_m[c,d]$ be proper $s$-tuples. With $m(H)$ we denote the set of all $m\in\N$ such that there exists $c,d\in\C$ for which
$H\mid\G_m[c\:d]$. And, the set of all proper $s$-tuples $H$ with infinite $m(H)$ will be denoted $\H^s$, or simply
$\H$ when it is clear from the context what $s$ is. We organize $H\in\H$ by the lowest
order $n$ at which $H$ divides a $\G_n$-tuple. By $\H_n$ we denote the set of all proper tuples $H$ with infinite
$m(H)$ whose smallest element is $n$.

We have the following lemma.
\begin{lemma} \label{lem1}
Equation (\ref{ceveq}) is nonlinear injective and approximately
integrable iff there is a proper 6-tuple $H$ with $m(H)$ infinite,
such that $\G_n[a\:b]$ divides $\widehat{K}^1H$.
\end{lemma}
\proof
\begin{itemize}
\item[$\Leftarrow$] The fact that $\G_n[a\:b]$ divides a proper tuple implies that equation (\ref{ceveq}) is nonlinear
injective. The equation is approximately integrable because for every $m\in m(H)$ there are $c,d$ such that
\[
\widehat{S}^1=\frac{\G_m[c\:d]}{H}\frac{\widehat{K}^1H}{\G_n[a\:b]}
\]
is proper.
\item[$\Rightarrow$]
Because equation (\ref{ceveq}) is nonlinear injective, the tuple $\widehat{S}^1=\G_m[c\:d]\widehat{K}^1/\G_n[a\:b]$ is
well defined for all $m$. The integrability implies that $S^1$ is proper for infinitely many $m\in\N$ and $c,d\in\C$.
This only happens when $\G_n=HP$ factorizes such that $P\mid \widehat{K}^1$ and $m(H)$ is infinite.
\qed
\end{itemize}

According to Lemma \ref{lem1}, to classify approximately integrable nonlinear injective equations it suffices to
determine the set $\H^6$ of all proper 6-tuples $H$ with infinite $m(H)$.  This will be done using results from number
theory, provided in section \ref{numthe}. In section \ref{1} we determine
the proper divisors $H\in\H^1$ of infinitely many functions $\G^{i,1-i}_{k,m}$ for possible $i,k$. And, in section \ref{2}
we determine the proper divisors $H\in\H^2$ of infinitely many 2-tuples $\G^{i,1-i}_{1,m}$, $\G^{j,1-j}_{k,m}$,
where $i\neq j$ if $k=1$. From those results we determine the set $\H^6=\bigcup_{n\in\N}\H^6_n$ in section \ref{6}.
For each $n\in\N$ the set $\H_n$ is related to the set of approximate integrable equations at order $n$, which are
not in a lower order hierarchy.

We would like to provide an explicit but minimal list of approximate integrable equations from which one can derive
all approximately integrable equations. The following observation is useful. Let $P$ and $Q$ be proper tuples. From
Lemma \ref{lem1} it follows that if equation (\ref{ceveq}), with $\widehat{K}^1=P$, is approximately integrable, then
the same equation, but with $\widehat{K}^1=PQ$, is also approximately integrable. Therefore, the classification in
section \ref{6} describes the divisors that have maximal degree. And the corresponding list of equations only contains
equations with quadratic parts $K^1$ of minimal degree.

From the results of sections \ref{1}, \ref{2} it follows that $\H_n$ is non-empty for all $n\in\N$.
That means there are new approximately integrable equations at every order. In section \ref{6} we
classify the highest degree divisors in $\H_n$ completely, that is, for any order. We are not able to
explicitly list all corresponding equations, as this paper is bound to be finite. In section \ref{6} we
do provide a complete list of approximately integrable equations of order $n \leq 5$.

We explicitly provide the linear parts $(cu_m,du_m)$ of all the symmetries of the equations in our list.
This enables one to calculate any approximate symmetry in principle. This can be done using Maple code provided
at \cite{MC}. We remark that if one multiplies the quadratic tuple of an equation with a proper tuple,
the resulting equation may have more symmetries than the original one. As we will now illustrate it may also be in a
lower hierarchy.

From Lemma \ref{lem1} we know that if $H\in\H_n$ and $\G_n[a\:b]\mid \widehat{K}^1H$, then equation (\ref{ceveq}) is
approximately integrable with approximate symmetries at (higher) order $m\in m(H)$. The following lemma applies.
\begin{lemma} \label{lem2}
Suppose $H\in\H_n$ and $\G_n[a\:b]$ divides $\widehat{K}^1H$. Then equation (\ref{ceveq}) has more symmetries
than the ones at order $m\in m(H)$ iff there is a divisor $Q\in\H_{k\leq n}$ of $H$, with $m(H)$ smaller than
and contained in $m(Q)$, such that $\G_n[a\:b]$ divides $\widehat{K}^1Q$.
\end{lemma}
\proof Given a divisor $Q\in\H_k$ of $H$ such that $\G_n[a\:b]\mid \widehat{K}^1Q$, it is clear that equation
(\ref{ceveq}) has a symmetry at every order $m\in m(Q)$ with
\[
\widehat{S}^1=\frac{\G_m[c\:d]}{Q}\frac{\widehat{K}^1Q}{\G_n[a\:b]}.
\]
To see that the converse holds, let $Y$ denote the set of orders of approximate symmetries, with $m(H)$ smaller than
and contained in $Y$. We need to prove that there is a $Q$ such that $Y=m(Q)$. Take $m\in Y\setminus m(H)$ and
write $\G_n=HP$. Since $\G_n\mid \widehat{K}^1H$ we have $\widehat{K}^1=PR$. The tuple
$\widehat{S}^1=\G_m\widehat{K}^1/\G_n=\G_mR/H$ is proper. Since $m\not\in m(H)$, $H$ does not divide
$\G_m$. There is a proper divisor $Q$ of $H$ such that $Q\mid \G_m$ and $H/Q$ divides $R$, that is,
$\G_n\mid \widehat{K}^1Q$. Since $Q\mid H$ the set $m(Q)$ is infinite.
\qed

\begin{remark} \label{rem}
One can start with an equation that is not nonlinear injective, multiply its quadratic tuple, and end
up in the hierarchy of an nonlinear injective equation. For example, apart from certain purely nonlinear symmetries,
equation 1.2 has approximately symmetries with linear part $(cu_m,dv_m)$ for any $c,d\in\C$ when $m$ is odd.
By multiplying its quadratic tuple with the tuple $[0,(f_1x_1+f_2y_1)/f,(y_1+y_2)/2,
0,(i_1x_1+i_2y_1)/i,(x_1+x_2)/2]$ we obtain the equation
\[
\vec{u_t}{v_t}=\vec{au_1+f_1u_1v+f_2uv_1+gvv_1}{v_1+i_1u_1v+i_2uv_1+juu_1},
\]
which has approximate symmetries at all orders $m>0$ for any $c,d\in\C$, and, it is in the hierarchy of
an equation of the form 0.3 iff $f_1=i_2=0$. In this paper we do not explicitly describe all symmetries of all
approximately integrable equations that can be obtained from our list.
\end{remark}

\section{Results from number theory} \label{numthe}
Generally speaking, progress in classifying global classes of evolution equations
has been going hand in hand with applying new results or techniques
from number theory. For the classification of scalar equations \cite{MR99g:35058}
the new result was obtained by F. Beukers, who applied sophisticated techniques from
diophantine approximation theory \cite{MR98e:11029}. The Skolem--Mahler--Lech theorem,
stated below, first appeared in the literature in connection with symmetries of evolution equations in
\cite{MR99i:35005}. Beukers, Sanders and Wang used a partial corollary of this theorem to conjecture that
there are only finitely many integrable equations (\ref{ceveq}) with $K^1=[0,0,1,0,0,0]$.
Their conjecture became a theorem in \cite{MR1829636}, where an exhaustive
list of the integrable cases was produced using a recent algorithm of C.J. Smyth \cite{BS}, that
solves polynomial equations $f(x,y)=0$ for roots of unity $x,y$. And, the classification of $\B$-equations
was due to results on diophantine equations in roots of unity, again proved by Beukers \cite{PH02}.

However, as it turns out, we do not need entirely different results or techniques from number theory
to globally classify two component evolution equations, with homogeneous diagonal linear part, admitting
infinitely many approximate symmetries.

\subsection{The Skolem--Mahler--Lech theorem} \label{sml}
A sequence $\{U_m, m\in\N\}$ satisfies an order $n$ linear
recurrence relation if there exist $s_1,\ldots,s_n$ such that
\[
U_{m+n}=s_1U_{m+n-1}+\cdots s_nU_m.
\]
The general solution can be expressed in terms of a generalized
power sum
\[
U_m=\sum_{i=1}^k A_i(m)\alpha_i^m,
\]
such that the roots $\alpha_i$ are distinct and non-zero, and
the coefficients $A_i(m)$ are polynomial in $m$. By definition
the degree of $U_m$ is $d=\sum_{i=1}^k d_i$, where $d_i$ is the
degree of $A_i(m)$. It can be shown that the order of the sequence
equals $n=k+d$ \cite{ajvdp}.\footnote{In \cite{ajvdp} one should
replace equation 2.1.2 by equation 1.3 from \cite{zors}.}

A generalized power sum vanishes identically, $U_m=0$ for
all $m$, precisely when all its coefficients vanish as polynomials in
$m$, $A_i(m)=0$ for all $i$. We prove this by induction on the
degree. For $d=0$ the statement is plain, the functions $h\rightarrow
\alpha_i^h$ are linearly independent for distinct $\alpha_i$. Let
$S:f(m)\rightarrow f(m+1)$ be the shift operator. Suppose $d>0$. Then
for some $i$ we have $d_i>0$. The generalized power sum
$V_m=(S-\alpha_i)U_m$ has degree $d-1$. By the induction hypothesis
we have, in particular, $\alpha_i(S-1)A_i(h)=0$. Since $\alpha_i\neq
0$ this implies $d_i=0$ and hence we are done.

\begin{theorem}[Skolem--Mahler--Lech] \label{smlt}
The zero set of a linear recurrence sequence $\{m\in\N : U_m=0\}$ is
the union of a finite set and finitely many complete arithmetic
progressions.
\end{theorem}

Note that an arithmetic progression $p$ is complete if $p=\{f+gh:h\in\N\}$
for some {\em remainder} $f\in\N_0$ and {\em difference} $g>f$, $g\in\N$.
Theorem \ref{smlt} was first proved by Skolem for the rational numbers
\cite{Sko}, by Mahler for algebraic numbers \cite{Mah}, and by Lech
for arbitrary fields of characteristic zero \cite{Le53}. The proofs
rely on $p$-adic analysis and consist of showing the existence of a
difference $g\in \N$ such that every partial sum, with $0\leq f< g$,
\begin{equation} \label{pas}
U_{f+gh}=\sum_{i=1}^k (A_k(f+gh)\alpha_i^f)(\alpha_i^{g})^h
\end{equation}
either has finitely many solutions $h$ or vanishes identically. We
refer to \cite{myer,vesa}, and references in there, for sensible
sketches of a proof.

If (\ref{pas}) vanishes identically the sum on the right breaks up
into disjoint pieces $I\subset \{1,\ldots, m\}$ each of which
vanishes because the roots $\alpha_i^{g}$, $i\in I$, coincide and the
sum of their coefficients $\sum_{i\in I} A_i(f+gh)\alpha_i^f$
vanishes identically as a function of the variable $h$. Since
$A_i(f+gh)$ does not vanish identically, each piece contains at least
two terms. In particular, the following will be usefull.

\begin{corollary} \label{smlc}
If the equation
\[
a_1\alpha_1^m + a_2\alpha_2^m + \cdots + a_k\alpha_k^m = 0,
\]
with nonzero $a_i,\alpha_i\in\C$ has infinitely many solutions,
the set $\{\alpha_1,\alpha_2,\ldots,\alpha_k\}$ partitions into
a number of disjoint subsets, such that each subset has at least
two members, and the ratio of any two members of a subset is a
root of unity.
\end{corollary}
For instance, when $k=3$ the triple $\alpha_1/\alpha_2, \alpha_2/\alpha_3,
\alpha_1/\alpha_3$ consists of roots of unity.

\subsection{Diophantine equations in roots of unity}
The following theorems are of crucial importance for
the classification problem considered in this paper.

\begin{theorem}[Beukers] \label{diop1}
Take $m>1$ integer. Let \(\mu,\nu\) be distinct roots of unity, both
not equal to 1, such that $\nu\neq\mu^{-1}$ when $m$ is odd. Then
\begin{equation} \label{de1}
(1-\nu^m)(1-\mu)^m=(1-\mu^m)(1-\nu)^m
\end{equation}
implies $\mu^m=\nu^m=1$.
\end{theorem}

\begin{theorem}[Beukers] \label{diop2}
Take $m>1$ integer. Let \(\mu,\nu\) be distinct roots of unity, not
both equal to 1, such that $\nu\neq\mu^{-1}$ when $m$ is even. Then
\begin{equation} \label{de2}
(1+\nu^m)(1-\mu)^m=(1+\mu^m)(1-\nu)^m
\end{equation}
implies $\mu^m=\nu^m=-1$.
\end{theorem}

\begin{theorem}[Beukers] \label{diop3}
Take $m>1$ integer. Let \(\mu,\nu\) be roots of unity with $\mu\neq
1$. Then
\begin{equation} \label{de3}
(1+\nu^m)(1-\mu)^m=(1- \mu^m)(1-\nu)^m
\end{equation}
implies $\mu^m=-\nu^m=1$.
\end{theorem}

Whereas the Skolem--Mahler--Lech theorem implies that certain ratios
are roots of unity for the equation to have infinitely many solutions,
the above theorems tell us precisely what the solutions are. In particular,
they imply that the zero sets consist of complete arithmetic progressions only.

Theorems \ref{diop1},\ref{diop2},\ref{diop3} are slightly more general than \cite[Theorems 22,25]{PH02},
which were proved by F. Beukers. We won't repeat their proofs here, however we do indicate the difference
between the two sets of Theorems, which is twofold. Firstly, in Theorems \ref{diop1},\ref{diop2},\ref{diop3} we do not
assume that $\mu,\nu\neq -1$. In certain cases this follows from \cite[Proposition 24]{PH02}, in others one
has to rely on the following.
\begin{proposition}[Beukers] \label{prop}
If $\nu$ is a root of unity such that
\begin{equation} \label{ntr}
(1+\nu^m)2^{m-1}=(1-\nu)^m,
\end{equation}
then $\nu=-1$ and $m$ is even.
\end{proposition}
\proof By Galois theory we may assume that $\nu=e^{2\pi i/n}$. Taking $n=1$ does not
give any solutions. If $n=2$ then $m$ has to be even. We will show there are no solutions
with $n>2$. When $m=1$ there is no root of unity such that $1+\nu=1-\nu$. Taking $m=2$ it
follows that $n=2$. So we may assume that $m>2$.

Since $\nu\neq 1$, $|1+\nu^m|$ does not vanish and we have $|1+\nu^m|>\sin(\pi/n)$.
Also we use $|1-\nu|<2\pi/n$. This gives
\[
(2\frac{\pi}{n})^m>|1-\nu|^m=|1+\nu^m|2^{m-1}>\sin(\frac{\pi}{n})2^{m-1}.
\]
Division by $2^m\pi/n$ yields (taking $n>2$)
\[
(\frac{\pi}{n})^{m-1} > \sin(\frac{\pi}{n}) \frac{n}{2\pi} > .41,
\]
which implies (taking $m>2$) that $\pi/n > .64$, or $n<5$.
When $n=3$, $|1+\nu^m|$ equals 1 or 2, and $|1-\nu|=\sqrt{3}$, whose $m$-th power
does not equal $2^m$ or $2^{m-1}$. When $n=4$, $|1+\nu^m|$ equals 0 or $\sqrt{2}$ or $2$, and
$|1-\nu|=\sqrt{2}$, whose $m$-th power, with $m>1$, does not equal $0$ or $\sqrt{2}2^{m-1}$
or $2^m$.
\qed

Secondly in the proofs we do not in general need $\nu\neq\mu$ and $\nu\neq 1/\mu$.
And, we also note that in \cite[Theorem 25]{PH02} it was mistakenly supposed
that $\mu^n\neq -1$. This should have been $\mu^n\neq \mp 1$ depending on
the sign in \cite[equation (10)]{PH02}.

\section{Homogeneous quadratic parts} \label{1}
In this section we determine the proper divisors of infinitely
many 1-tuples $\G_m=\G^{i,1-i}_{k,m}$ for all possible choices of $i,k$.

Due to equation (\ref{gequ}) we may take $k=1$; equations of the form $(u_t,v_t)=(au_n,bv_n+K)$
are related, by the linear transformation $u\leftrightarrow v$, to equations of the form
$(u_t,v_t)=(au_n+K,bv_n)$. We start with the simplest case $i=1$.

\subsection{Classifying approximately integrable scalar equations}
The Lie derivative of the quadratic part $S^1$ of a possible scalar symmetry
with respect to the linear part $K^0=u_n$ of a scalar equation
$u_t=K^0+K^1+\cdots$ is symbolically given by
$\L(K^0)S^1\trianglerighteq G^1_n \widehat{S}^1$ with
$\G$-function
\[
\G^1_n(x,y)=x^n+y^n-(x+y)^n=\G^{1,0}_{1;n}[a\:b](x,y)/a.
\]
Thus the case $i=k=1$ is equivalent to the scalar problem, which is easily seen by taking $v=0$.
The function is also proportional to $\G^{i,1-i}_{k,n}[a\:a]$ so the results
apply to the case $a=b$ as well.

In the classification of scalar equations \cite{MR99g:35058} a different route
was taken than the one we take. Namely, whereas we perform our classification with respect to the
existence of infinitely many (approximate) symmetries, Sanders and Wang performed their
classification with respect to the existence of symmetries (finitely many or infinitely many).
They showed in particular that there are no scalar equations with finitely many generalized
symmetries, which confirms the first part of the conjecture of
Fokas \cite{Fok87}:
\begin{quote}
If a scalar equation possesses at least one time-independent non-Lie
point symmetry,  then it possesses infinitely many. Similarly for
\(N\)-component equations one needs \(N\) symmetries.
\end{quote}
We note that the conjecture of Fokas does not hold inside the class of $\B$-equations \cite{KS99}.
In their classification Sanders and Wang relied on the following `hard to obtain' result
from number theory, proved in \cite{MR98e:11029}.
\begin{theorem}[Beukers] \label{da}
Let \(r\in\C\) such that \(r(r+1)(r^2+r+1)\neq0\). Then at most one
integer \(m>1\) exists such that \(\G^1_m(1,r)=0\).
\end{theorem}

In contrast, classifying the equations with respect to (approximate) integrability
can be done using the following `easy to obtain' result. Proposition \ref{110} is of course not as
strong as Theorem \ref{da}. For obvious reasons we do not include the constant divisors in $\H_0$ in our lists.
\begin{proposition} \label{110}
The proper divisors of infinitely many $\G^{1,0}_{1;m}[c\:d](1,y)$ are
products of
\begin{enumerate}
\item $y \in \H_2$, $m>1$
\item $(1+y)\in\H_3$, $m\equiv 1 \mod 2$
\item $1+y+y^2\in\H_5$, $m\equiv 1,5 \mod 6$
\item $(1+y+y^2)^2\in\H_7$, $m\equiv 1 \mod 6$
\end{enumerate}
\end{proposition}
\proof
According to the Skolem--Mahler--Lech theorem, see Corollary (\ref{smlc}), if the
diophantine equation \(\G^1_m(1,r)=0\) has infinitely many solutions
$m$, then $r=0,-1$ or $r$ and $r+1$ are both roots of unity, in which
case \(r\) is a primitive \(3\)-rd root of unity. The orders are
found by substituting the values for $r$. We have $\G^1_m(1,0)=0$ for
all $m$, $\G^1_{f+2h}(1,-1)=1+(-1)^f=0$ when $f=1$,
and, with $1+r+r^2=0$, $\G^1_{f+6h}(1,r)=1+r^f-(1+r)^f=0$ when $f=1$ or $f=5$.
Finally, by solving the simultaneous equations $\G^1_m(1,r)=\partial_r\G^1_m(1,r)=0$
we find that $r$ is a double zero when both $r$ and $1+r$ are $(m-1)$-st roots of
unity.
\qed

As a particular corollary of Proposition \ref{110} we have the following. Equation (\ref{ceveq})
with $a=b$ and $n=2,3,5,7$ is approximately integrable for arbitrary $K^1$.

\subsection{$\B$-equations}
The case $i=-1$ has been globally classified with respect
to integrability in \cite{PH02}. This class of equations is particularly nice
because any approximate symmetry is a symmetry. We go through the main ideas and
formulate the results slightly different from \cite{PH02}, minimizing the role of biunit coordinates.
This makes the argument cleaner and sets the stage for the main results of this paper. As the case
$c=d$ is covered in the previous section, this will be excluded in what follows.

\begin{proposition} \label{1-12}
All proper divisors $H$ of $\G^{-1,2}_{1;m}[c\:d](1,y)$ with $c\neq d$ and $m(H)$ infinite
can be obtained from the following list.
\begin{enumerate}
\item $1+y\in\H_1$, $m\equiv 1 \mod 2$, $d\neq 0$
\item $(1+y)^n \in\H_n$, $m\geq n$, $d=0$
\item $(y-r)(ry-1)\in\H_2$, $r\neq -1$, $m\geq 1$
\item $(y-r)^2(ry-1)^2\in\H_n$, $r\neq -1$, $n>3$ the smallest integer such that $r^{n-1}=1$,
$m \equiv 1\mod n-1$
\item $(y-r)(yr-1)(y-\bar{r})(y\bar{r}-1)\in\H_n$,
$r=\nu(\mu-1)/(\nu-1)$, $\mu,\nu$ roots of unity such that $(\mu-1)(\nu-1)(\mu-\nu)(\mu\nu-1)\neq 0$, $n>3$
the smallest integer such that $\mu^n=\nu^n=1$, $m \equiv 0 \mod n$
\item $1+y^n\in\H_n$, $m \equiv n \mod 2n$, $c=0$
\end{enumerate}
Unless stated otherwise, the coefficients of the linear part of the symmetries satisfy $c/d=(1+r^m)/(1+r)^m$.
\end{proposition}

\proof We study the zeros of the function
\[
\G^{-1,2}_{1;m}[c\:d](1,r)=d(1+r^n)-c(1+r)^n.
\]
Take $d\neq 0$. Then $r\neq-1$ is a zero when
\begin{equation} \label{ab}
\frac{c}{d}=\frac{1+r^m}{(1+r)^m},
\end{equation}
in which case $1/r$ is a zero as well. The point $r=-1$ is a zero
when $m$ is odd, where it has multiplicity 1, or when $d=0$, where
the multiplicity is $m$.

The other multiple zeros are obtained from
setting the $r$-derivatives of the function to zero, see also
\cite{MR99i:35005}. Taking $r\neq-1$ and solving the simultaneous
equations $\G^{-1,2}_{1;m}(1,r)=\partial_r\G^{-1,2}_{1;m}(1,r)=0$
yields $r^{m-1}=1$, while
$\partial_r\G^{-1,2}_{1;m}(1,r)=\partial_r^2\G^{-1,2}_{1;m}(1,r)=0$
yields $r=-1$. Therefore, all multiple zeros $r\neq-1$ are double
zeros. We have $c/d=1/(1+r)^{m-1}$ and $1/r$ is a double zero as well.
There are no other double zeros since the equations
$|r|=|s|$ and $|1+r|=|1+s|$ imply that $r=s$ or $r=\bar{s}$.
Let $n$ be the lowest integer such that $r^{n-1}=1$, so $r$ is a primitive
$(n-1)$-st root of unity. All $m$ such that $r^{m-1}=1$ are $m\equiv 1$ mod $n-1$.

To classify higher degree divisors we have to find all $r,s\in \C$, with
$(1+r)(1+s)(r-s)(rs-1)\neq 0$ such that the diophantine equation
\begin{eqnarray*}
U_m(r,s)&=&\G^{-1,2}_{1,m}[1+r^m,(1+r)^m](1,s)\\
&=&(1+r)^m+((1+r)s)^m-(1+s)^m-((1+s)r)^m = 0
\end{eqnarray*}
has infinitely many solutions $m$. According to the Skolem--Mahler--Lech theorem,
see Corollary \ref{smlc}, either $rs=0$ or one of the pairs
\begin{equation} \label{pairs1}
\frac{1+r}{1+s},\frac{(1+s)r}{(1+r)s} \quad \text{ or }\quad
\frac{1+r}{r(1+s)},\frac{1+s}{s(1+r)} \quad \text{ or }\quad r,s
\end{equation}
consists of roots of unity. When $rs=0$ we have $c=d$ which we exclude. Suppose the first pair of
(\ref{pairs1}) consists of roots of unity, let
$\mu=(1+r)/(1+s)$ and $\nu=(1+1/s)/(1+1/r)$.
We may write $r=\M(\mu,\nu)$, where
\[
\M(\mu,\nu)=\nu\frac{\mu-1}{\nu-1},
\]
and find that $s=\M(1/\mu,1/\nu)=\bar{r}$. In terms of roots of unity $\mu,\nu$ we have
\[
U_m(r,s)=\left(\frac{1-\mu\nu}{\mu(1-\nu)^2}\right)^m\left((1-\mu)^m(1-\nu^m)-(1-\nu)^m(1-\mu^m)\right).
\]
Note that $(\mu-1)(\nu-1)(\mu-\nu)(\mu\nu-1)\neq 0$ because $(r-s)(rs-1)\neq 0$.
Hence, using Theorem \ref{diop1}, we obtain $\mu^m=\nu^m=1$. When $n$ is the lowest integer such
that $\mu^n=\nu^n=1$, $\mu$ or $\nu$ are primitive $n$-th roots of unity. And all $m$ such that $\mu^m=\nu^m=1$
are given by $m\equiv 0 \mod n$. Next, suppose the second pair of (\ref{pairs1}) consists of roots of unity.
By a transformation $r\rightarrow 1/r$ we get the first
pair. Since $\M(\mu,\nu)^{-1}=\M(1/\nu,1/\mu)$ we get the same solutions, but with $s=1/\bar{r}$. Finally, when
$r,s$ are roots of unity $U_m(r,s)=0$ can be written in terms of $\mu=-r,\nu=-s$,
\[
U_m(r,s)=(1-\mu)^m(1+(-\nu)^m)-(1-\nu)^m(1+(-\mu)^m).
\]
When $m$ is odd Theorem \ref{diop1} applies and when $m$ is even Theorem \ref{diop2} applies.
\qed

\begin{figure}[hbt]
\begin{center}
\epsfig{file=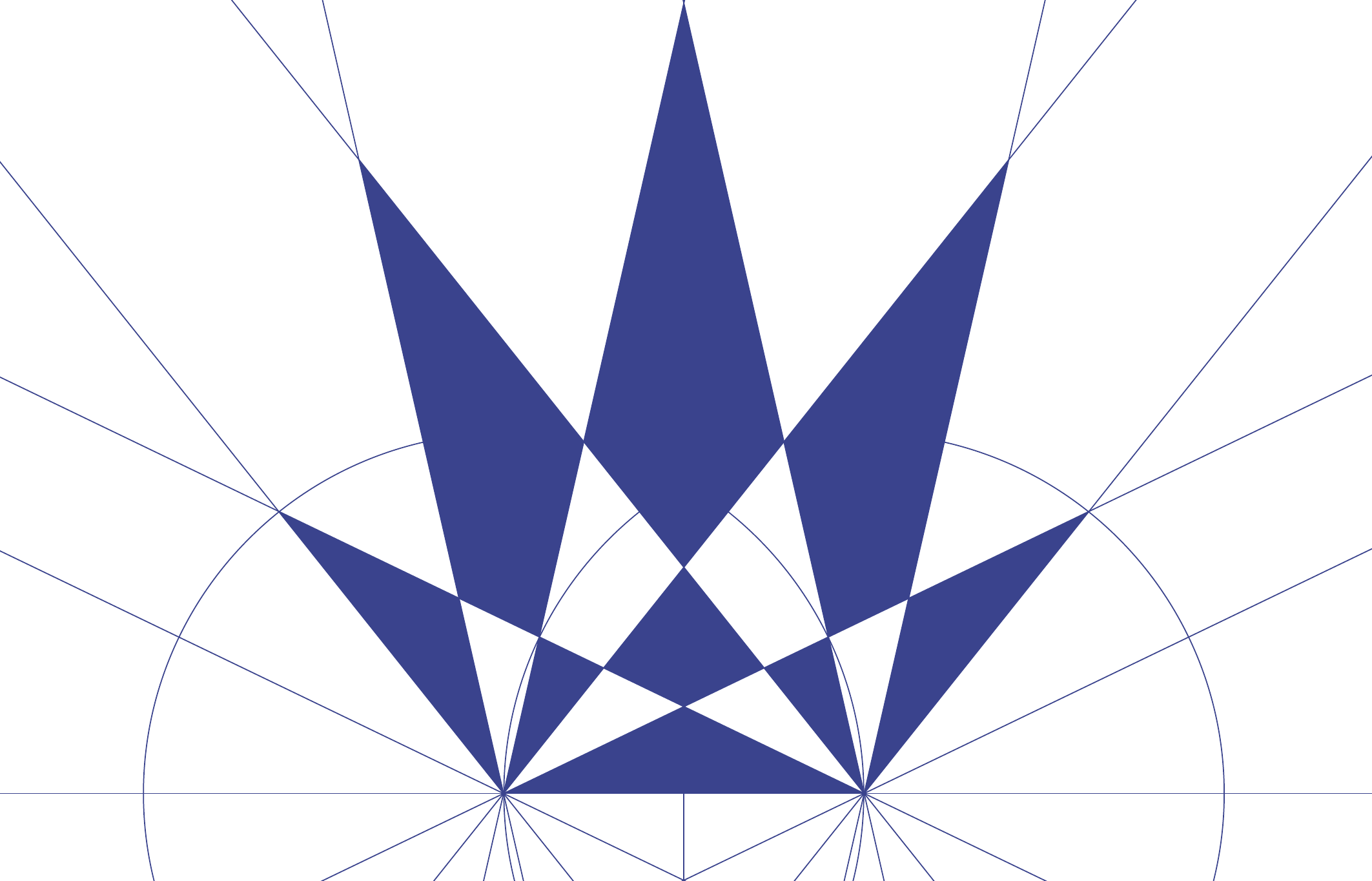, width = 8cm}
\caption{\label{patt} The corner
points $r$, with $|r| \neq 0,1$ satisfy $U_m(r,\bar{r})=0$
when $m\equiv 0$ mod $7$. The two circles are $|r|=1$ and $|r+1|=1$.}
\end{center}
\end{figure}

Consider the set of points
\begin{equation} \label{zet}
\{ r\in\C: r=\M(\mu,\nu), \mu^m=\nu^m=1, (\mu-1)(\nu-1)(\mu-\nu)(\mu\nu-1)\neq 0 \}.
\end{equation}
To illustrate where these points lie in the complex plane we use biunit coordinates.
Suppose $\psi,\phi$ are such that $|\psi|=|\phi|=1$ and $r$ is the unique intersection point
of the lines $\psi\R$ and $\phi\R-1$. Then $r=\r \psi\phi$, with
\[
\r \psi\phi = \psi^2\frac{\phi^2-1}{\psi^2-\phi^2},
\]
and $(\psi,\phi)$ are called the biunit coordinates of $r$. Denote further
\[
\r AB = \{r\in\C:r=\r ab, a\in A, b\in B, a^2 \neq b^2\},
\]
and
\[
\ROOT m= \{ r\in\C: r^m=1, r^2\neq 1\}.
\]
Using the algebraic relation $\M(\phi^2,\psi^2/\phi^2)=\r \psi\phi$ one verifies that the set (\ref{zet})
is equal to $\{r\in\r{\ROOT{2m}}{\ROOT{2m}}: \abs r \neq 1\}$. For $n=7$ the upper half of this set is plotted
in Figure \ref{patt}.

\subsection{Quadratic terms bilinear in $u$-, and $v$-derivatives}
This section deals with the case $i=0$.

\begin{proposition} \label{101}
If $H$ is a proper divisor of $\G^{0,1}_{1;m}[c\:d](1,y)$ with $m(H)$ infinite then $H$ is a product of
the following polynomials.
\begin{enumerate}
\item $y\in\H_1$, $m>0$, $c\neq 0$
\item $y^n\in\H_n$, $m\geq n$, $c=0$
\item $(y-r)\in\H_2$, $r\neq 0$, $m > 1$
\item $(y-r)(y(1+r)+r)\in\H_3$, $m \equiv 1\mod 2 $.
When $(1+r)^{2n}=1$, $r\neq 0$, we also have $m\equiv 0 \mod 2n$, $d=0$.
\item $(y-r)^2 \in \H_{2n}$, $r\neq 0$, $n$ the smallest integer such that $(1+r)^{2n-1}=1$,
$m \equiv 1\mod 2n-1$
\item $(y-r)^2(y(1+r)+r)^2 \in\H_{2n+1}$, $n>1$ the smallest integer such that $(1+r)^{2n}=1$,
$m \equiv 1\mod 2n$
\item  $(y-r)(y(1+r)+r)(y-\bar{r})(y(1+\bar{r})+\bar{r})\in\H_n$, $n>3$ odd, $r=(\mu-\nu)/(\nu-1)$,
$(\mu-1)(\nu-1)(\mu-\nu)(\mu\nu-1)\neq 0$, $n$ the smallest integer such that $\mu^n=\nu^n=1$,
$m \equiv n \mod 2n$
\item $(y-r)(y-\bar{r})\in\H_n$, $n>2$ even, $r=(\mu-\nu)/(\nu-1)$,
$(\mu-1)(\nu-1)(\mu-\nu)(\mu\nu-1)\neq 0$, $n$ the smallest integer such that $\mu^n=\nu^n=1$,
$m \equiv 0 \mod n$
\item $(y-r)(y(1+\bar{r})+\bar{r})\in\H_n$, $n>2$ even, $r=(\nu-\mu)/(\mu-1)$,
$(\mu-\nu)(\mu\nu-1)\neq 0$, $n$ the smallest integer for which $\mu^n=\nu^n=-1$,
$m \equiv n \mod 2n$
\end{enumerate}
Unless stated otherwise, the coefficients of the linear part of the symmetries satisfy $c/d=r^m/((1+r)^m-1)$.
\end{proposition}

\proof We are after the zeros of infinitely many
\[
\G^{0,1}_{1;m}[c\:d](1,y)=c-c(1+y)^m+dy^m.
\]
Take $c\neq0$. Then $r\neq 0$ is a zero of
precisely when
\begin{equation} \label{ba}
\frac{d}{c}=\frac{(1+r)^m-1}{r^m}.
\end{equation}
When $m$ is odd $-r/(1+r)$ is a zero as well. The point $r=0$ is a
zero for all $c,d,m$. It has multiplicity 1, except when $c=0$ where the multiplicity is $m$.
One can show that the multiple zeros $r\neq0$ of $\G^{0,1}_{1;m}[c\:d]$ are the double
zeros $\{r\neq 0: (1+r)^{m-1}=1\}$, with $c/d=r^{m-1}$. When $r$ is a
double zero the only other double zero is $\bar{r}= -r/(1+r)$ when $m$
is odd.

Higher degree divisors are given by distinct non-zero $r,s\in \C$, with
$r+rs+s\neq 0$ when $m$ is odd, such that the diophantine equation
\begin{eqnarray*}
U_m(r,s)&=&\G^{0,1}_{1,m}[r^m,(1+r)^m-1](1,s)\\
&=& r^m-r^m(1+s)^m+s^m(1+r)^m-s^m=0
\end{eqnarray*}
has infinitely many solutions $m$. The
cases $r=-1$, $s=-1$ yield the primitive third roots of unity, as in Proposition \ref{110},
where $c=d=1$, which we excluded.
Then, according to the the Skolem--Mahler--Lech theorem, at least one
of the pairs
\begin{equation} \label{pairs2}
\frac{r}{s},\frac{r(1+s)}{s(1+r)} \quad \text{ or }\quad
\frac{s}{r(1+s)},\frac{s}{r}(1+r)\quad \text{ or }\quad 1+r,1+s
\end{equation}
consists of roots of unity. Suppose the first pair consist of roots
of unity. Let $\mu=r/s$ and $\nu=r(1+s)/s/(1+r)$. Then $(\mu-1)(\nu-1)(\mu-\nu)\neq 0$,
$\mu\nu\neq 1$ when $m$ odd, $r={\cal N}(\mu,\nu)$ and $s={\cal N}(1/\mu,1/\nu)=\bar{r}$ with
\[
{\cal N}(\mu,\nu)=\frac{\mu-\nu}{\nu-1}.
\]
When $\mu\nu=1$ and $m$ even we have $\bar{r}=-r/(1+r)$.
In terms of $\mu,\nu$ we get
\[
U_m(r,s)=\left(\frac{\nu-\mu}{\mu(\nu-1)^2}\right)^m
\left((1-\mu)^m(1-\nu^m)-(1-\nu)^m(1-\mu^m)\right),
\]
which implies, using Theorem \ref{diop1}, that $\mu^m=\nu^m=1$.
 In bi-unit coordinates we have $r\in
\r{\ROOT{2m}}{\ROOT{2m}}$ such that $\abs{r+1}\neq 1$ when $m$ odd.

Next, suppose that the second pair of (\ref{pairs2}) consists of
roots of unity, $\mu=-r/s/(1+r)$, $\nu=-(1+s)r/s$. We have
$(\mu-1)(\nu-1)(\mu-\nu)(\mu\nu-1)\neq 0$ when $r+rs+s\neq0$, that
is, when $m$ odd. When $r+rs+s=0$ and $m$ even we get $(1+r)^m=1$,
which corresponds to $b=0$. Otherwise, $r={\cal K}(\mu,\nu)=(\nu-\mu)/(\mu-1)$
and $s=-\bar{r}/(1+\bar{r})$. In terms of $\mu,\nu$ we have
\[
U_m(r,s)=\left(\frac{\nu-\mu}{\mu(\mu-1)(\nu-1)}\right)^m
\left((1-\nu)^m(1+(-\mu)^m)-(1-\mu)^m(1+(-\nu)^m)\right)
\]
When $m$ is odd Theorem \ref{diop1} implies $\mu^m=\nu^m=1$, while
for $m$ even Theorem \ref{diop2} yields $\mu^m=\nu^m=-1$. The biunit
coordinate description can be found as follows. Solve the
simultaneous equations ${\cal K}(\mu,\nu)=\r \psi\phi$, ${\cal
K}(1/\mu,1/\nu)=\r {1/\psi}{1/\phi}$ to find that
$\mu=\psi^2/\phi^2,\nu=\psi^2$. For odd $m$ we don't find new values
for $r$, but for $m$ even we get
$r\in\r{\ROOT{4m}\setminus\ROOT{2m}}{\ROOT{2m}}$ such that
$|r+1|\neq 1$. Finally, suppose that the last pair of (\ref{pairs2})
consists of roots of unity. Then $\mu=1+r$ and $\nu=1+s$ satisfy
equation (\ref{de1}). According to Theorem \ref{diop1} we have
$(1+r)^m=(1+s)^m=1$, that is, the second eigenvalue equals 0.
\qed

Actually, when $m$ is odd the two cases $i=-1,i=0$ are related. We have
\begin{equation} \label{rel}
\G^{0,1}_{1;m}[c\:d](1,r)=\G^{-1,2}_{1;m}[c\:d](1,-1-r).
\end{equation}
Indeed, at odd order $m$ the zero $r=-1$ of $\G^{-1,2}_{1,m}$
translates into the zero $r=0$ of $\G^{0,1}_{1,m}$. Also the image of
the unit circle $|z|=1$ under $f_3:r\rightarrow -1-r$ is the unit
circle $|z+1|=1$, relating the double zeros of the two $\G$-functions.
The symmetry $f_2: r\rightarrow 1/r$ is translated into
$f_4=f_3 \circ f_2 \circ f_3: r\rightarrow -r/(1+r)$.
And we note that set $\r{\ROOT{m}}{\ROOT{m}}$ is invariant under the
group of an-harmonic ratios, generated by $f_2$ and $f_3$, cf. \cite{MR98g:14032}.
Using the above, for odd $m$ one may obtain Propostion \ref{101} from Proposition
\ref{1-12} and vise versa.

Summarizing this section, it implies that equations with homogeneous qua-dratic
parts are approximately integrable when $n<4$. At any order $n\geq 4$ a finite number
of new approximately integrable equations has been found.

\section{Non-homogeneous quadratic parts} \label{2}
This section deals with equations whose quadratic part is not homogeneous, that is,
$K^1=K^{i,1-i}_1,K^{j,1-j}_k$ with $i\neq j$ when $k=1$. We provide the corresponding
sets $\H^2_n$ of $2$-tuples. This time we do find conditions on the ratio $a/b$ for
low orders $n<4$.

When $i=1$ the first part of the condition $H\in\H^2_n$, $H_{[1]}\in\H^1$ being a divisor of
infinitely many $\G^{1,0}_{1;m}$, does not give conditions on $c/d$, see Proposition
\ref{110}. In this case the $\H^2_n$ are obtained from the classification
of $H_{[2]}\in\H^1$ dividing infinitely many $\G^{j,1-j}_{k;m}$, which was obtained
in the previous section. A similar remark can be made when $(j,k)=(0,2)$.
Due to equation (\ref{gequ}) there are four cases left to consider, with
$k=1$: $(i,j)=(-1,0)$; and with $k=2$: $(i,j)=(0,1)$, $(i,j)=(-1,2)$, $(i,j)=(-1,1)$.

There are certain divisors of infinitely many $\G_m[c\:d]$-functions
for any value of $c/d$. These will be called trivial divisors. Apart from
the constant divisors we have
\begin{eqnarray*}
(1+y) \mid \G^{-1,2}_{1;2m+1}(1,y) & & y\mid \G^{-1,2}_{1;m}(1,y) \\
(x+1) \mid \G^{2,-1}_{2;2m+1}(x,1) & & x\mid \G^{1,0}_{2;m}(x,1)
\end{eqnarray*}
We may take $H_{[1]}$ (or $H_{[2]}$) to be trivial. Then $H\in\H^2$ if $H_{[2]}$ ($H_{[1]}$) is one of
the divisors of infinitely many $\G$-functions presented in the previous section.
In the sequel we assume that neither $H_{[1]}$ nor $H_{[2]}$ is trivial. Also we will assume that
$cd(c-d)\neq 0$.

\begin{proposition} \label{1-121}
We list the non-trivial divisors $H$ of the 2-tuple $\G^{-1,2}_{1;m}[c\:d](1,y)$,
$\G^{1,0}_{2;m}[c\:d](x,1)$ with $m(H)$ infinite. Firstly suppose $n$ is odd and
$P(y)$ divides $\G^{-1,2}_{1;m}[c\:d](1,y)$ with infinite $m(P)$ whose smallest element is $n$,
cf. Proposition \ref{1-12}. Then $P(y),P(-1-x) \in \H_n$. Secondly, when $n$ is even
we have:
\begin{enumerate}
\item $(y-r)(ry-1),x+1\in\H_2$, $r\in\PROOT{3}$, $m\equiv 2,4 \mod 6$
\item $(y-r)^2(ry-1)^2,x+1\in\H_4$, $r\in\PROOT{3}$, $m\equiv 4 \mod 6$
\item $(y-r)(ry-1),\bar{r}x+\bar{r}+1  \in \H_n$, $r= -\nu(\mu-1)/\mu/(\nu-1)$, $\mu\neq 1$,
$n$ the lowest integer such that $\mu^n=-\nu^n=1$, $m\equiv n \mod 2n$.
\end{enumerate}
The linear coefficients of the symmetries satisfy $c/d=(1+r^m)/(1+r)^m$.
\end{proposition}

\proof
When the order of the equation $n$ is odd, no new conditions on the
linear part are obtained since the relations (\ref{gequ}) and
(\ref{rel}) imply that
\[
\G^{-1,2}_{1;m}[c\:d](1,-1-r)=\G^{1,0}_{2;m}[c\:d](r,1).
\]

For even $n$, there should be $r\in\C$ with $s \neq 0$ such that
\[
\G^{-1,2}_{1;m}[c\:d](1,r)=\G^{1,0}_{2;m}[c\:d](s,1)=0,
\]
or, equivalently,
\begin{equation} \label{U}
U_m(r,s)=s^m+(rs)^m+(1+r)^m-((1+r)(1+s))^m=0
\end{equation}
for infinitely many $m$ including $n$. Then, using the
Skolem-Mahler-Lech theorem, we may infer that either $rs(1+r)(1+s)=0$
or at least one of the pairs
\begin{equation} \label{pairs3}
r,1+s\qquad \frac{rs}{1+r},\frac{s}{(1+s)(1+r)},\qquad
\frac{rs}{(1+r)(1+s)},\frac{s}{1+r}
\end{equation}
consists of roots of unity. When $r(r+1)=0$ we have $c=d$ or $d=0$, which we excluded.
When $s=-1$ we are left with the equation $U_m=(-1)^m+(-r)^m+(1+r)^m=0$.
Applying the Skolem--Mahler--Lech theorem we see that both $r$ and $1+r$ are
roots of unity and hence, that $r$ is a primitive third root of unity. One
verifies that $U_{i+3k}=\left((-1)^i+(-r)^i+(1+r)^i\right)(-1)^k=0$
when $i$ equals 1 or 2. Also, if $-1$ is a zero of
$\G^{1,0}_{2;m}[c\:d]$, then $c/d=-(-1)^m$.

Suppose the first pair of (\ref{pairs3}) consists of roots of unity. Writing
equation (\ref{U}) in terms of $\mu=1+s$, $\nu=-r$, we get equation (\ref{de3}).
Theorem \ref{diop3} then implies $\mu^m=-\nu^m=1$, which corresponds to the case
$a=0$, which we excluded. Suppose the second pair of (\ref{pairs3}) consists of roots
of unity. Then $\mu = s/(1+s)/(1+r)$ and $\nu = -rs/(1+r)$ are roots of unity, and
we get $r = {\cal K}(u,v) = -\nu(\mu-1)/\mu/(\nu-1)$, $s=-(1+\bar{r})/\bar{r}$, and
\[
U_m=\left(\frac{\mu-\nu}{(\mu-1)(\nu-1)\mu}\right)^m((1+(-\nu)^m)(1-\mu)^m-(1-\mu^m)(1-nu)^m).
\]
When $m$ is even, Theorem \ref{diop3} yields $\mu^m=-\nu^m=1$ or $\mu=1$. But, when $\mu=1$ we have
$s=-(1+r)/r$ and $U_{m}=2(1+r)^{m}=0$ iff $r=-1$, which we excluded. Using ${\cal K}(1/\phi^2,\psi^2/\phi^2)=\r{\psi}{\phi}$ we may write
$r\in\r{\ROOT{4m}\setminus\ROOT{2m}}{\ROOT{2m}}$.
When $n$ is even and $n$ is the lowest integer such that $\mu^n=-\nu^n=1$ we have $\mu$ is a primitive
$n$-th root of unity or $\nu$ is a primitive $2n$-th root of unity and all solutions to $\mu^n=-\nu^n=1$
are given by $m\equiv n \mod 2n$.

The third pair of (\ref{pairs3}) is obtained from the second by $f_2:r\rightarrow 1/r$. Under this transformation
we have $\r \psi\phi \rightarrow \r{\psi^{-1}}{\phi\psi^{-1}}$. Hence we get the solutions
$r\in\r{\ROOT{4m}\setminus\ROOT{2m}}{\ROOT{4m}\setminus\ROOT{2m}}$ and $s=-1-\bar{r}$. Or one can express
$U_m=0$ in terms of $\mu = rs/(1+r)/(1+s)$, $\nu = -(1+r)/s$ to find these values. Another way of
describing the last item would be: {\it 3. $[(y-r)(ry-1),x+\bar{r}+1]\in\H_n$, $r= \mu(\nu-1)/(\mu-1)$, $\mu\neq 1$,
$n$ the lowest integer such that $\mu^n=-\nu^n=1$, $m\equiv n \mod 2n$}.
\qed

In the remaining cases the diophantine equation we obtain
from the zeros of the $\G$-functions will be of the form
\begin{equation} \label{abc}
(1+aA^m)(1+bB^m)+cC^m=0.
\end{equation}
\begin{lemma} \label{abcl}
Suppose that the diophantine equation (\ref{abc}), with $ABC\neq 0$, has infinitely many solutions.
Then $A$, $B$, and $C$ are roots of unity.
\end{lemma}
\proof
Using Corollary \ref{smlc}, three of the numbers $1,A,B,AB,C$ have a root of
unity as a ratio and the same is true for the remaining two. Therefore at least
one of the pairs $C,A$; $C,B$; $C/A,B$; $C/B,A$
consists of roots of unity. When $C$ and $A$ are roots of unity, their
powers yield a finite number of values. Moreover, for the infinite number
of solutions we have $(1+aA^m)\neq 0$. Hence, for these infinite number
of solutions $(1+bB^m)$ has only finitely many values. This only happens
when $B$ is a root of unity. The other cases lead to the same result, e.g.
when $C/A$ and $B$ are roots of unity we divide the equation by $A^m$ and
find that $A$ is a root of unity.
\qed

Suppose that the triple $\zeta,\eta,f(\zeta,\eta)$ consist of roots of unity. Then we
can apply the algorithm of Smyth \cite{BS} to solve the equation
$
f(\zeta,\eta)^{-1}=f(\zeta^{-1},\eta^{-1})
$
for roots of unity. In particular, a finite number of values will be obtained.
We denote the set of all primitive $n$-th roots of unity by \(\PROOT{n}\).

\begin{proposition} \label{1-110}
We list the non-trivial divisors $H$ of the tuple $\G^{-1,2}_{1;m}[c\:d](1,y)$,
$\G^{0,1}_{1;m}(1,y)$ with $m(H)$ infinite.
\begin{enumerate}
\item
$y^2+y+1,y+1\in\H_{2}$, $m\equiv 1,2 \mod 3$, $c/d=-(-1)^m$
\item
$(y^2+y+1)^2,y+1\in\H_{4}$, $m\equiv 1 \mod 3$, $c/d=-(-1)^m$
\item
$(y-r^2)(y-\bar{r}^2)$, $(y-r+1)(y-\bar{r}+1)\in\H_3$, $r \in \PROOT{10}$,
$m \equiv 1,3,7,9 \mod 10$
\item
$(y-r^2)^2(y-\bar{r}^2)^2$, $(y-r+1)^2(y-\bar{r}+1)^2\in\H_{11}$, $r \in \PROOT{10}$,
$m \equiv 1 \mod 10$
\item
$(y-r)(y-\bar{r})$, $(y-r+1)(y-\bar{r}+1)\in\H_5$, $r \in \PROOT{12}$,
$m \equiv 1,5,7,11 \mod 12$
\item
$(y-r)^2(y-\bar{r})^2$, $(y-r+1)^2(y-\bar{r}+1)^2\in\H_{13}$, $r \in \PROOT{12}$,
$m \equiv 1\mod 12$
\end{enumerate}
The linear coefficients of the symmetries satisfy $c/d=(r-1)^m/(r^m-1)$.
\end{proposition}

\proof
We have $\G^{-1,2}_{1;m}[c,d](1,r)=\G^{0,1}_{1;m}[c,d](1,s)=0$ when
\begin{equation} \label{U2}
U_m(r,s)=(1+r^m)(1-(1+s)^m)+(s(1+r))^m=0.
\end{equation}
We want to classify all $r,s\in \C$, with $rs(1+r)\neq 0$, such that
equation (\ref{U2}) has infinitely many solutions. According to Lemma
\ref{abcl} we have $s=-1$, or $r,1+s,s(1+r)$ consists of roots of unity.
When $s=-1$ we obtain that $r$ is a third root of unity and $U_{i+3k}=0$
iff $i=1,2$. When $x=r,y=1+s,f=s(1+r)$ consists of roots of unity, then
$x,y$ are cyclotomic points on the curve
\[
1+(xy-2(x-y))(xy-1)+(x-y)^2=0,
\]
and can be found algorithmically. They are $x\in\PROOT{3}$, $y\in\PROOT{6}$;
$y\in\PROOT{10}$, $x=y^2$ or $x=\bar{y}^2$; $y\in\PROOT{12}$, $x=y$ or $x=\bar{y}$.
The first case only happens when $c=d$. In the second case we have
$f=y^4$ or $f=y^2$ and we find $U_{i+10k}=0$ iff $i\in\{1,3,7,9\}$.
Note that with $s=y-1$, $\abs{y}=1$ we have $-s/(1+s)=\bar{s}$.
The last case gives $f=y^4$ or $f=y^3$ and $U_{i+12k}=0$ iff $i\in\{1,5,7,11\}$.

The multiplicity of the zeros is obtained from Proposition \ref{1-12} and \ref{101}.
We have $r\in\PROOT{3}$ is a double zero of $\G^{-1,2}_{1;m}$ when
$m\equiv 1$ mod $3$. When $y\in\PROOT{10}$ we have that both $r=y^2$ and $r=\bar{y}^2$
are in $\PROOT{5}$. They are double zeros of $\G^{-1,2}_{1;m}$ for $m\equiv 1$ mod $5$.
Also, we have that $s=y-1$ and $\bar{s}$ are double zeros of $\G^{0,1}_{1;m}$ for
$m\equiv 1$ mod $10$. A similar argument shows the multiplicity in the last item.
\qed
\begin{proposition} \label{1-122}
We list the non-trivial divisors $H$ of the tuple $\G^{-1,2}_{1;m}[c\:d](1,y)$,
$\G^{2,-1}_{2;m}(x,1)$ with $m(H)$ infinite:
\begin{enumerate}
\item
$(y-r)(y-\bar{r}),(x-r)(x-\bar{r})\in\H_2$, $r\in\PROOT{3}$,
$m\equiv 1,2 \mod 3$
\item
$(y-r)^2(y-\bar{r})^2,(x-r)^2(x-\bar{r})^2\in\H_4$, $r\in\PROOT{3}$,
$m\equiv 1 \mod 3$
\item
$(y-r^2)(y-\bar{r}^2)$, $(x-r)(x-\bar{r})\in\H_3$, $r \in \PROOT{5}$,
$m \equiv 1,3,7,9 \mod 10$
\item
$(y-r^2)^2(y-\bar{r}^2)^2$, $(x-r)^2(x-\bar{r})^2\in\H_{11}$, $r \in \PROOT{5}$,
$m \equiv 1 \mod 10$
\item
$(y+r)(y+\bar{r})$, $(x-r)(x-\bar{r})\in\H_4$, $r \in \PROOT{12}$,
$m \equiv 1,4,5,7,8,11 \mod 12$
\item
$(y+r)^2(y+\bar{r})^2$, $(x-r)^2(x-\bar{r})^2\in\H_{13}$, $r \in \PROOT{12}$,
$m \equiv 1\mod 12$
\end{enumerate}
The linear coefficients of the symmetries satisfy $c/d=(r+1)^m/(r^m+1)$.
\end{proposition}
\proof
We have $\G^{-1,2}_{1;m}[c,d](1,r)=\G^{2,-1}_{2;m}[c,d](s,1)=0$ when
\begin{equation} \label{U3}
U_m(r,s)=(1+r^m)(1+s^m)-((1+s)(1+r))^m=0.
\end{equation}
We want to classify all $r,s\in \C$, with $rs(s+1)(1+r)\neq 0$, such that
equation (\ref{U3}) has infinitely many solutions. According to Lemma
\ref{abcl}, the points $r$, $s$, and $(1+s)(1+r)$ are roots of unity.
Hence $r,s$ are cyclotomic points on the curve
\[
1+(rs+1)(rs+2(r+s))+(r+s)^2=0.
\]
Smyths algorithm yields: $r,s\in\PROOT{3}$; $s\in\PROOT{5}$, $r=s^2$ or $r=\bar{s}^2$;
$s\in\PROOT{12}$, $r=-s$ or $r=-\bar{s}$. Substitution these into the equation (\ref{U3}),
we obtained, by performing some Groebner basis calculations, the solutions $m\equiv 1,2\mod 3$,
$m\equiv 1,3,7,9\mod 10$, and $m\equiv 1,4,5,7,8,11\mod 12$ respectively. The multiplicities are
determined using Proposition \ref{1-12}, and using relation (\ref{gequ}).
\qed

\begin{proposition} \label{1021}
We list the non-trivial divisors $H$ of the tuple $\G^{0,1}_{1;m}[c\:d](1,y)$,
$\G^{1,0}_{2;m}(x,1)$ with $m(H)$ infinite:
\begin{enumerate}
\item
$y+1-r,x+1-r\in\H_2$, $r=0$, $m>1$
\item
$(y+1+r^2)(y+1+\bar{r}^2)$, $(x+1-r)(x+1-\bar{r})\in\H_3$, $r \in \PROOT{10}$,
$m \equiv 1,3,7,9 \mod 10$
\item
$(y+1+r^2)^2(y+1+\bar{r}^2)^2$, $(x+1-r)^2(x+1-\bar{r})^2\in\H_{11}$, $r \in \PROOT{10}$,
$m \equiv 1 \mod 10$
\item
$(y+1+r)$, $(x+1-r)\in\H_2$, $r \in \PROOT{12}$,
$m \equiv 1,2,5,7,10,11 \mod 12$
\item
$(y+1+r)(y+1+\bar{r})$, $(x+1-r)(x+1-\bar{r})\in\H_5$, $r \in \PROOT{12}$,
$m \equiv 1,5,7,11 \mod 12$
\item
$(y+1+r)^2(y+1+\bar{r})^2$, $(x+1-r)^2(x+1-\bar{r})^2\in\H_{13}$, $r \in \PROOT{12}$,
$m \equiv 1\mod 12$
\end{enumerate}
The linear coefficients of the symmetries satisfy $c/d=(r^m-1)/(r-1)^m$.
\end{proposition}
\proof Similar to the above, $\G^{0,1}_{1;m}[c,d](1,r)=\G^{1,0}_{2;m}[c,d](s,1)=0$ when
\begin{equation} \label{U4}
(1-(1+r)^m)(1-(1+s)^m)-(rs)^m=0.
\end{equation}
We want to classify all $r,s\in \C$, with $rs\neq 0$, such that
equation (\ref{U4}) has infinitely many solutions. If one of $r,s$ equals $-1$, the other
is a third root of unity. When $r=s=-1$ we have $a/b=-(-1)^m$, otherwise $c=d$.
Suppose that $(1+r)(1+s)\neq 0$. According to Lemma \ref{abcl} $1+r$, $1+s$, and $sr$ are roots of unity.
Then $x=1+r,y=1+s$ are cyclotomic points on the curve
\[
1+(xy+1)(xy-2(x+y))+(x+y)^2=0.
\]
They are: $x,y\in\PROOT{6}$; $y\in\PROOT{10}$, $x=-y^2$ or $x=-\bar{y}^2$;
$y\in\PROOT{12}$, $x=-y$ or $x=-\bar{y}$. The first are zeros only when $c=d$
and the others yield the results.
\qed

\section{Global classification of maximal degree divisors} \label{6}
Combining the results obtained in Propositions \ref{110}, \ref{1-12}, \ref{101}, \ref{1-121},
\ref{1-110}, \ref{1-122}, \ref{1021} we determine the set of all highest degree proper 6-tuples with
infinite $m(H)$. This is equivalent to a global classification of approximately integrable
two-component equations with a diagonal linear part, see section \ref{nsc}. Highest degree tuples are
formed as follows. With $H\in\H_n$ and $F\in\H_k$ we have $HF\in \H_l$, where $l$ is the smallest
number in $m(HF)=m(H)\cap m(F)$.

Clearly if $H$ divides $\G_m[c,d]$ then $H$ divides $\G_m[c/d,1]$. Also, if
\[
H=[O(x_1,x_2),P(x_1,y_1),Q(y_1,y_2),R(y_1,y_2),S(x_1,y_1),T(x_1,x_2)]
\]
divides $\G_m[c,1]$ then, according to equation (\ref{gequ}), the function $\G_m[1/c,1]$
admits the proper divisor
\[
H^\dagger=[R(x_1,x_2),S(y_1,x_1),T(y_1,y_2),O(y_1,y_2),P(y_1,x_1),Q(x_1,x_2)].
\]
Thus we scale $d$ in $\G_m[c,d]$ to 1, and perform the classification up to inversion of $c$.

We also include tuples with zero components in the list. They correspond to equations that are not
nonlinear injective, see section \ref{ni}. For each $K^0$ in Table \ref{tab} we have determined the highest
degree $r$-tuple $H$, with $m(H)$ infinite, which divides the $r$-tuple consisting of the non-zero
components of its $\G_n$-tuple. The quadratic tuple $K^1$ has $r$ non-zero components $K^1=\G_n/H$,
unless complementary components of the $\G_m$-tuple vanish at infinitely $m\in m(H)$.

First we deal with $\H^6_{n\leq5}$. Here we translate our symbolic results into differential language.
For any $H\in\H_n$ of highest degree, which divides $\G_n[a,1]$, we determine $K^1$ from
$\widehat{K}^1=[e,f,g,h,i,j]\G_n[a,1]/H$. In principal, the tuple $[e,f,g,h,i,j]$ may consist of
proper polynomials, it is a common factor of $\widehat{K}^1$ and $\widehat{S^1}$. However, when writing down
the differential equation, the $e,f,g,h,i,j$ will appear in it as constants, and any other constants will be
absorbed by them. This organizes the quadratic part of the equations. At the same time it may remind the reader
of the fact that the quadratic tuple of the equations can be multiplied by arbitrary proper tuples. We give each
maximal degree tuple $H\in\H_n$ two indices, $H=H_{n.h}$, where $n$ is the order, and $h$ is a counter. We label
the corresponding approximately integrable equation by $n.h$. And, we express the linear coefficients $c/d$ of
the approximate symmetries in terms of integer sequences, or in its power sum solution if that displays well.

Secondly we give a general description of $\H^6_{n>5}$. Using this result one can, in principle,
write down the corresponding approximate integrable systems at any particular order. Some Maple
code has been provided at \cite{MC}.

\setcounter{subsection}{-1}
\subsection{Zeroth order}
According to Table \ref{tab} there are two special values of $a$ related to equations that are
not nonlinear injective. We have the following 6-tuples in $\H_0$. At $a=0$
\[
H_{0.1} = [0 , 1, 1, 1, 0^\ast, 1].
\]
Both the first and fifth component of the zeroth order $\G$-tuple vanishes. However, there exist
higher order $\G$-tuple with zero first component, but no higher order $\G$-tuples with zero fifth
component. This is denoted by the ${}^\ast$, which indicates that in the approximate integrable equation
the term $K^{1,0}_2$ vanishes.
The equation
\begin{equation} \tag*{0.1}
\vec{u_t}{v_t}=\vec{eu^2+fuv+gv^2}{v+hv^2+ju^2}
\end{equation}
has approximate symmetries at order $m=1$, for all $c/d\in\C$, and at any order $m$, with $c=0$.

At $a=2$ we have
\[
H_{0.2} = [1, 1, 0^\ast, 1, 1, 1].
\]
The equation
\begin{equation} \tag*{0.2}
\vec{u_t}{v_t}=\vec{2u+eu^2+fuv}{v+hv^2+iuv+ju^2}
\end{equation}
has approximate symmetries at any order $m\in\N$, for all $c,d\in\C$.

And, at generic values of $a$ we have
\[
H_{0.3} = [1, 1, 1, 1, 1, 1].
\]
The equation, with $a(2a-1)(a-2)\neq0$,
\begin{equation} \tag*{0.3}
\vec{u_t}{v_t}=\vec{au+eu^2+fuv+gv^2}{v+hv^2+iuv+ju^2}
\end{equation}
has approximate symmetries at any order $m\in\N$, for all $c,d\in\C$.

\subsection{First order}
We have the following 6-tuples in $\H_1$. At $a=1$
\[
H_{1.1} = [0, 0^\ast, 0^\ast, 0^\ast, 0^\ast, 0^\ast].
\]
The equation
\begin{equation} \tag*{1.1}
\vec{u_t}{v_t}=\vec{u_1+eu^2}{v_1}
\end{equation}
has approximate symmetries at any order $m\in\N$, for $c=0$.
Of course, when $e=0$ any $S\in\g$ is a symmetry of this equation.

At generic values of $a$ we have
\[
H_{1.2} = [0, y_1, y_1+y_2, 0^\ast, x_1, x_1+x_2].
\]
The equation, with $a\neq 1$,
\begin{equation} \tag*{1.2}
\vec{u_t}{v_t}=\vec{au_1+eu^2+fuv+gv^2}{v_1+iuv+ju^2}
\end{equation}
has approximate symmetries at odd orders $m\equiv 1\mod 2$ for $c=0$.
Again, when $e=0$ there are more approximate symmetries, namely
at odd orders $m\equiv 1\mod 2$ for all $c,d\in\C$, see Remark \ref{rem}.

\subsection{Second order}
At second order there is the tuple
\[
H_2 = [x_1x_2, y_1, 1, y_1y_2, x_1, 1]
\]
which divides all higher $\G_m[c,d]$ for all $c,d\in\C$. The
maximal degree divisors $H_{2.i}\in\H_2$ are $H_{2.i}=H_2T_{2.i}$ with
\begin{eqnarray*}
T_{2.1} &=& [0, y_1, y_1^2+y_2^2, 1, x_1+2y_1, (x_1+x_2)^2] \\
T_{2.2} &=& [1, x_1, y_1y_2, 1, y_1, x_1x_2] \\
T_{2.3} &=& [1, x_1+y_1, y_1^2+y_1y_2+y_2^2, 1, x_1+y_1, x_1^2+x_1x_2+x_2^2] \\
T_{2.4} &=& [1, 1, (ry_2-y_1)(ry_1-y_2), 1, 1, 1] \\
T_{2.5} &=& [1, rx_1-y_1, 1, 1, 1, 1] \\
T_{2.6} &=& [1, 1, (y_1+y_2-\iota y_2)(2y_1+y_2+y_2 \iota), 1, 2x_1+y_1+y_1\iota , 1] \\
T_{2.7} &=& [1, 2y_1-\iota x_1+2x_1+x_1\gamma, 1, 1, 2x_1-y_1\gamma -\iota y_1+2y_1,1],
\end{eqnarray*}
where $\iota^2=-1,\gamma^3=3$.

The equation
\begin{equation} \tag*{2.1}
\vec{u_t}{v_t}=\vec{eu^2+fuv+gv^2}{v_2+hv^2+iuv+ju^2}
\end{equation}
has approximate symmetries at orders $m\equiv 2 \mod 4$ with $c=0$.
The equation
\begin{equation} \tag*{2.2}
\vec{u_t}{v_t}=\vec{u_2+eu^2+fuv+gv^2}{v_2+hv^2+iuv+ju^2}
\end{equation}
has approximate symmetries at all orders $m>1$ with $c=d=1$. The equation
\begin{equation} \tag*{2.3}
\vec{u_t}{v_t}=\vec{-u_2+eu^2+fuv+gv^2}{v_2+hv^2+iuv+ju^2}
\end{equation}
has approximate symmetries at order $m\equiv 1,2 \mod 3$, with $c/d=-(-1)^m$.
The equation, with $(1+r^2)(1+r)\neq 0$,
\begin{equation} \tag*{2.4}
\vec{u_t}{v_t}=\vec{\frac{1+r^2}{(1+r)^2}u_2+eu^2+ f(ruv_1 - (1+r^2)u_1v)+ gv^2}{v_2
+hv^2+ i (ru_1v-(1+r)^2uv_1) + j(2ru_2u+(1+r)^2u_1^2)}
\end{equation}
has approximate symmetries at all orders $m>1$ with $c/d=(1+r^m)/(1+r)^m$.
The equation, with $r(r+2)\neq 0$,
\begin{equation} \tag*{2.5}
\vec{u_t}{v_t}=\vec{\frac{r}{2+r} u_2+eu^2+fuv+g(rv_1^2-2vv_2)}{v_2
+hv^2+i(u_1v+(2+r)uv_1)+j(2uu_2+(2+r)u_1^2)}
\end{equation}
has approximate symmetries at all orders $m>1$ with $c/d=r^m/((1+r)^m-1)$. 

The equation, with $\iota^2=-1$,
\begin{equation} \tag*{2.6}
\vec{u_t}{v_t}=\vec{(-1+2\iota)u_2+eu^2+f(5u_1v+(3+\iota)uv_1)+gv^2}{v_2
+hv^2+iuv+j(4uu_2+(1+\iota)u_1^2)}
\end{equation}
has approximate symmetries, with $c/d=-1+(-1)^{(m-2)/4} 2^{m/2} \iota$, at order $m\equiv 2 \mod 4$.
The equation, with $\gamma^2=3$, $\iota^2=-1$,
\begin{equation} \tag*{2.7}
\vec{u_t}{v_t}=\vec{\iota(2+\gamma)u_2 +eu^2+fuv+g(4vv_2+(2+\gamma-\iota)v_1^2)}{v_2
+hv^2+iuv+j(4uu_2+(2+\iota-\gamma)u_1^2)}
\end{equation}
has approximate symmetries at orders $m \equiv q \mod 12$, with $q\in\{1,2,5,7,10,11\}$. Define
integers $P_k$ by $P_1=1$, $P_2=2$, and
\begin{equation} \label{P}
P_k=\left\{ \begin{array}{ll} P_{k-1}+P_{k-3} & k\equiv 1 \mod 3 \\
P_{k-1}+P_{k-2} & k\equiv 0,2 \mod 3 \end{array} \right. .
\end{equation}
When $q = 2$ or $q=10$ the coefficients of the linear part of the approximate symmetries of equation 2.7 are
given by
\[
\frac{c}{d}=(-1)^{(m-q)/12}\iota(P_{3m/2-1}+P_{3m/2-2}\gamma),
\]
or else by
\begin{equation} \label{cd}
\mp (-1)^{(m-q)/12} \frac{c}{d}=\left\{ \begin{array}{ll}
P_{(3m-5)/2} + P_{(3m-7)/2}\gamma & q = 6 \pm 5 \\
P_{(3m+1)/2} + P_{(3m-1)/2}\gamma & q = 6 \pm 1
\end{array} \right. .
\end{equation}
The sequence $\{P_k\}$ is quite interesting in itself, see \cite[Sequence A140827]{Slo}. It satisfies the 6-th
order recurrence $P_k=4P_{k-3}-P_{k-6}$. And thus it consists of the three subsequences
\cite[Sequences A001075, A001353, A001835]{Slo}. The first two of these subsequences are the denominators
and numerators of convergents to $\sqrt{3}$, we have $P_{3n-1}^2-3P_{3n-2}^2=1$.

\subsection{Third order}
At third order the product
\[
H_3 = H_2[x_1+x_2, 1, y_1+y_2, y_1+y_2, 1, x_1+x_2]
\]
divides $\G_m[c,d]$ when $m$ odd, for all $c,d\in\C$.
The maximal degree divisors $H_{3.i}\in\H_3$ are $H_{3.i}=H_3T_{3.i}$ where
\begin{eqnarray*}
T_{3.1} &=& [0, y_1^2, y_2^2-y_1y_2+y_1^2, 1, x_1^2+3x_1y_1+3y_1^2, (x_1+x_2)^2] \\
T_{3.2} &=& [1, x_1(x_1+y_1), y_1y_2, 1, y_1(x_1+y_1), x_1x_2] \\
T_{3.3} &=& [1, 1, (y_1-ry_2)(y_1r-y_2), 1, (rx_1+(1+r)y_1)(x_1+(1+r)y_1), 1] \\
T_{3.4} &=& [1, x_1^2+x_1y_1+(2-\phi)y_1^2, y_1^2+\phi y_1y_2+y_2^2, 1, x_1^2+(2-\phi)(x_1y_1+y_1^2),\\
&& \hspace{7.2cm}
x_1^2+(1-\phi) x_1x_2 + x_2^2],
\end{eqnarray*}
in which $\phi$ denotes the golden ratio or its conjugate, that is, $\phi(\phi-1)=1$.
Note that both $H_{2.4}$ and $H_{2.5}^\dagger|_{-1-r}$ divide $H_{3.3}$.

The equation
\begin{equation} \tag*{3.1}
\vec{u_t}{v_t}=\vec{eu^2+fuv+gv^2}{v_3+hv^2+iuv+ju^2}
\end{equation}
has approximate symmetries at orders $m\equiv 3 \mod 6$ with $c=0$.
The equation
\begin{equation} \tag*{3.2}
\vec{u_t}{v_t}=\vec{u_3+eu^2+fuv+gv^2}{v_3+hv^2+iuv+ju^2}
\end{equation}
has approximate symmetries at odd orders with $c=d=1$. The equation, with $r^3\neq-1$,
\begin{equation} \tag*{3.3}
\vec{u_t}{v_t}=\vec{\frac{1-r+r^2}{(1+r)^2}u_3+eu^2+f(ruv_2-(1-r+r^2)(u_2v+u_1v_1)}{v_3
+hv^2+iuv+j(2ruu_2+(1+r+r^2)u_1^2)}
\end{equation}
has approximate symmetries at odd orders $m$ with $c/d=(1+r^m)/(1+r)^m$.
The equation, with $\phi(\phi-1)=1$,
\begin{equation} \tag*{3.4}
\vec{u_t}{v_t}=\vec{-(2+3\phi)u_3+eu^2+fuv+gv^2}{v_3+hv^2+iuv+ju^2}
\end{equation}
has approximate symmetries at order $m\equiv q \mod 10$, $q\in\{1,3,7,9\}$, with
\begin{equation} \label{fib}
c/d=\left\{ \begin{array}{ll}
F_{m-2}+F_{m-1}\phi & q=5\pm 4 \\
-F_m-F_{m+1}\phi & q=5\pm 2 \\
\end{array} \right. ,
\end{equation}
where the $F_k$ are the Fibonacci numbers $F_0=0$, $F_1=1$, $F_k=F_{k-1}+F_{k-2}$
\cite[Sequence A000045]{Slo}.

\subsection{Fourth order}
At order four the maximal degree divisors $H_{4.i}\in\H_4$ are $H_{4.i}=H_2T_{4.i}$ where
\begin{eqnarray*}
T_{4.1} &=& [0, y_1^3, y_1^4+y_2^4, 1, (x_1+2y_1)(x_1^2+2x_1y_1+2y_1^2), (x_1+x_2)^4] \\
T_{4.2} &=& [1, x_1+y_1, (y_1^2+y_1y_2+y_2^2)^2, 1, x_1+y_1, (x_1^2+x_1x_2+x_2^2)^2] \\
T_{4.3} &=& [1, 1, (y_1^2+2y_1y_2+2y_2^2)(2y_1^2+2y_1y_2+y_2^2), 1, 1, 1] \\
T_{4.4} &=& [1, (2y_1+(3-\gamma \iota)x_1)^2, 1, 1, 1, 1] \\
T_{4.5} &=& [1, x_1^2+2x_1y_1+2y_1^2, 1, 1, 1, 1] \\
T_{4.6} &=& [1, y_1^2-2\iota x_1y_1-(1+\iota)x_1^2, 1, 1, 1, 1] \\
T_{4.7} &=& [1, y_1^2+2(1-\iota\beta )x_1y_1-2\iota\beta x_1^2, 1, 1, 1, 1] \\
T_{4.8} &=& [1, 1, 4y_1^2+(6-2\iota+2\iota\beta )y_1y_2+4y_2^2, 1,
2x_1+(1-\iota+\iota\beta )y_1, 1] \\
T_{4.9} &=& [1, 1, 2y_1^2+(1-\iota-\iota\beta -2\beta )y_1y_2+2y_2^2, 1,
2x_1 + (1 + \iota + \beta )y_1, 1] \\
T_{4.10} &=& [1, 1, 4y_1^2+(6-4\iota-3\iota\beta -\beta )y_1y_2+4y_2^2, 1,
(1+\iota-\iota\beta )x_1 + y_1, 1] \\
T_{4.11} &=& [1, 1, y_1^2+y_1y_2\gamma +y_2^2, 1, 1, x_1^2-x_1x_2\gamma +x_2^2],
\end{eqnarray*}
with $\iota^2=-1,\beta^2=2,\gamma^3=3$.

The equation
\begin{equation} \tag*{4.1}
\vec{u_t}{v_t}=\vec{eu^2+fuv+gv^2}{v_4+h(4vv_2+3v_1^2)+iuv+ju^2}
\end{equation}
has approximate symmetries at orders $m\equiv 4 \mod 8$ with $c=0$.
The equation
\begin{equation} \tag*{4.2}
\vec{u_t}{v_t}=\vec{-u_4+e(4u_2u+3u_1^2)+f(uv_2+u_1v_1+2u_2v)+gv^2}{v_4
+h(4v_2v+3v_1^2)+i(2uv_2+u_1v_1+u_2v)+ju^2}
\end{equation}
has approximate symmetries at order $m\equiv 1 \mod 3$ with $c/d=-(-1)^m$.
The equation
\begin{equation} \tag*{4.3}
\vec{u_t}{v_t}=\vec{\begin{split}-3u_4+e(4u_2u+3u_1^2)+f(6u_3v+9u_2v_1+6u_1v_2 \\ +2uv_3) +gv^2 \end{split}}{
\begin{split}v_4 + h(4v_2v+3v_1^2)+i(2u_3v+2u_2v_1+3u_1v_2+2uv_3) \\ +j(4u_4u+4u_1u_3+3u_2^2) \end{split} }
\end{equation}
has approximate symmetries at order $m\equiv 0 \mod 4$ with $c/d=1+(-1)^{m/4}2^{m/2}$. 
The equation, with $\zeta^2+\zeta+1=0$,
\begin{equation} \tag*{4.4}
\vec{u_t}{v_t}=\vec{\begin{split}3(1+2\zeta)u_4+e(4uu_2+3u_1^2)+f(6u_1v+(1-4\zeta)uv_1) \\
+g(14v_4v+3(4-\zeta)(4v_3v_1+3v_2^2))\end{split}}{
\begin{split} v_4 +h(4v_2v+3v_1^2)+i(7u_3v+(2+3\zeta)(2u_2v_1+2uv_3\\
+3u_1v_2))+j(14u_4u+(2+3\zeta)(4u_3u_1+3u_2^2))\end{split}}
\end{equation}
has approximate symmetries at order $m\equiv 1 \mod 3$, with
\[
\frac{c}{d} = \left\{ \begin{array}{ll}
(-3)^{(m-1)/2}, & m\equiv 1 \mod 6, \\
-(1+2\zeta)(-3)^{(m-2)/2}, & m\equiv 4 \mod 6.
\end{array} \right.
\]
The equation, with $\epsilon=\pm 1$,
\begin{equation} \tag*{4.5}
\vec{u_t}{v_t}=\vec{\begin{split} \epsilon/5 u_4 + e(4uu_2+3u_1^2) + f(2u_1v+(1-2\epsilon)uv_1) \\
+ g( 4 v_3v_1 + 3 v_2^2 + (1-5\epsilon) v_4v)\end{split}}{
\begin{split} v_4 + h(4v_2v+3v_1^2) + i(10(2u_2v_1+3u_1v_2+2uv_3) \\+(5-\epsilon) u_3v)
+ j (5(4u_3u_1+3u_2^2)+(5-\epsilon)u_4u)\end{split}}
\end{equation}
has approximate symmetries at order $m\equiv 0 \mod 4$, with
\[
\frac{c}{d}=\frac{\epsilon}{1-(-4)^{m/4}}.
\]
The equation, with $\iota^2=-1$,
\begin{equation} \tag*{4.6}
\vec{u_t}{v_t}=\vec{\begin{split} \iota/3 u_4 + e(4uu_2+3u_1^2) + f(2u_1v + (1 - 2 \iota) uv_1) \\
+g(10vv_4+(1-3\iota)(4v_1v_3+3v_2^2))\end{split}}{
\begin{split} v_4 + h(4v_2v+3v_1^2) + i(5u_3v+3(3+\iota)(2u_2v_1 + 3 u_1v_2 \\+ 2uv_3))
+ j(10u_4v+3(3+\iota)(3u_2^2+4u_3u_1)) \end{split}}
\end{equation}
has approximate symmetries at orders $m=4+k8$, $k\in\N$, with
\[
\frac{c}{d}=\frac{(-1)^k2^{2k}\iota}{2A_{k+1}^2 + 1},
\]
where the integers $A_i$ are the
NSW numbers defined by $A_0=-1$ ,$A_1=1$, $A_i=6A_{i-1}-A_{n-2}$ \cite[Sequence A002315]{Slo}. 
The equation, with $\alpha^2=-2$
\begin{equation} \tag*{4.7}
\vec{u_t}{v_t}=\vec{\begin{split} \alpha   u_4 + e(4uu_2+3u_1^2)  +
f(2u_1v+(1-\alpha  )uv_1) \\ +g(2(4v_1v_3+3v_2^2)+(2+\alpha  )v_4v)\end{split}}{
\begin{split} v_4 +h(4v_2v+3v_1^2) + i((1-\alpha  )u_3v+2(2u_2v_1+3u_1v_2 \\+2uv_3))
 +j(3u_2^2+4u_3u_1+(1-\alpha  )u_4v) \end{split}}
\end{equation}
has approximate symmetries at orders $m=4+k8$, $k\in\N$, with
\[
\frac{c}{d}=\frac{(-1)^k2^{6k}\alpha  }{B_{k+1}},
\]
where the integers $B_i$ are defined
by $B_0=-1$ ,$B_1=1$, $B_i=34B_{i-1}-B_{n-2}$ \cite[Sequence A046176]{Slo}. 
The equations, with $\iota^2=-1,\beta^2=2$,
\begin{equation} \tag*{4.8}
\vec{u_t}{v_t}=\vec{\begin{split} (-1+2\iota (3+\beta ))u_4 + e(4uu_2+3u_1^2)  +
f(6uv_3 \\ +(9-\iota+2\beta )(2u_1v_2+3u_2v_1+2u_3v_0)) \\
+g(12vv_2+(9+\beta +\iota)v_1^2)\end{split}}{
\begin{split} v_4 +h(4v_2v+3v_1^2) + i(6u_2v+(3+5\iota-3\iota\beta -4\beta )u_1v_1  \\ +2(1+3\iota-\iota\beta )uv_2 )
+j(12uu_4 \\ +(3+\iota-2\beta )(3u_2^2+4u_3u_1)) \end{split}}
\end{equation}
\begin{equation} \tag*{4.9}
\vec{u_t}{v_t}=\vec{\begin{split} (-17+12\beta +2\iota (3-2\beta ))u_4 + e(4uu_2+3u_1^2)  +
f(10uv_3 \\ +(11-3\iota-6\beta -2\iota \beta )(2u_1v_2+3u_2v_1+2u_3v_0)) \\
+g(20vv_2+(17-2\beta -\iota+\iota\beta )v_1^2)\end{split}}{
\begin{split} v_4 +h(4v_2v+3v_1^2) + i(10u_2v+(13+\iota+7\beta +4\beta \iota)u_1v_1  \\ +2(7-\iota+5\beta )uv_2 )
+j(20uu_4 \\ +(9+3\iota+2(3+\iota)\beta )(3u_2^2+4u_3u_1)) \end{split}}
\end{equation}
\begin{equation} \tag*{4.10}
\vec{u_t}{v_t}=\vec{\begin{split} (17-12\beta +8\iota (3-2\beta ))u_4 + e(4uu_2+3u_1^2)  +
f(12uv_3 \\ -(3\beta +4\iota \beta -20+6\iota )(2u_1v_2+3u_2v_1+2u_3v_0)) \\
+g(24vv_2+(22-3\beta +\iota\beta )v_1^2)\end{split}}{
\begin{split} v_4 +h(4v_2v+3v_1^2) + i(2uv_2+(1+2\iota(\beta  -1))u_1v_1  \\ +(\iota-5+(4-\iota)\beta )u_2v )
+j(24uu_4 \\ +(4+6\iota+(3+4\iota)\beta )(3u_2^2+4u_3u_1)) \end{split}}
\end{equation}
have approximate symmetries at orders $m=4+k8$, $k\in\N$, with
\begin{eqnarray*}
\frac{c}{d}&=&-1+\iota(-1)^{k}2^{2k+1}(C_{m+1}+C_m\beta ),\\
\frac{c}{d}&=&(-1+\iota(-1)^{k}2^{2k+1}(C_{m+1}+C_m\beta ))(C_{2m+1}-C_{2m}\beta ),\\
\frac{c}{d}&=&(1+\iota(-1)^{k}2^{6k+3}(C_{m+1}+C_m\beta ))(C_{2m+1}-C_{2m}\beta ),
\end{eqnarray*}
respectively, where $C_0=0$, $C_1=1$, and $C_{2n}=C_{2n-1}+C_{2n-2}$, $C_{2n+1}=2C_{2n}-C_{2n-1}$.
These integers, cf. \cite[Sequence A002965]{Slo}, are the denominators and numerators of convergents
to $\sqrt{2}$, we have
\[
C_n^2-2C_{n-1}^2=\pm 1,\ \text{ when }\ n\equiv 1\pm 1 \text{ mod } 4.
\]
The equation, with $\gamma^2=3$
\begin{equation} \tag*{4.11}
\vec{u_t}{v_t}=\vec{\begin{split} (7+4\gamma)u_4 + e(4uu_2+3u_1^2)+ f(2u_1v_2 +3u_2v_1+2u_3v \\ + (2\gamma-3)uv_3)
+g(6vv_2+(6+\gamma)v_1^2) \end{split}}{
\begin{split} v_4 + h(4v_2v+3v_1^2) + i((3+2\gamma)u_3v-2uv_3-3u_1v_2\\-2u_2v_1)
+ j(6uu_2+(6-\gamma)u_1^2) \end{split} }
\end{equation}
has approximate symmetries at order $m\equiv q\mod 12$, $q\in \{1,4,5,7,8,11\}$. When $q = 6\pm 2$
the coefficients of the linear part of the approximate symmetries are given by
\[
\frac{c}{d}=\mp (-1)^{(m-q)/12}(P_{3m/2-1}+P_{3m/2-2}\gamma),
\]
where the integers $P_k$
are defined by the recursive formula (\ref{P}). When $q$ is odd, $c/d$ is given by equation
(\ref{cd}).

\subsection{Fifth order}
At order five the maximal degree divisors $H_{5.i}\in\H_5$ are $H_{5.i}=H_5T_{5.i}$ where
\[
H_5=H_3[x_1^2+x_1x_2+x_2^2,1,1,y_1^2+y_1y_2+y_2^2,1,1]
\]
and
\begin{eqnarray*}
T_{5.1} &=& [0, y_1^4, y_1^4-y_2y_1^3+y_1^2y_2^2-y_2^3y_1+y_2^4, 1, x_1^4+5x_1^3y_1+10x_1^2y_1^2+10x_1y_1^3 \\
&&\hspace{8cm} +5y_1^4, (x_1+x_2)^4] \\
T_{5.2} &=& [1, 2y_1^2x_1^2+2x_1^3y_1+x_1^4+y_1^3x_1, y_1^3y_2+y_1^2y_2^2+y_2^3y_1, 1, 2y_1^3x_1+2y_1^2x_1^2 \\
&&\hspace{5.5cm} +x_1^3y_1+y_1^4, x_1^3x_2+x_1^2x_2^2+x_1x_2^3] \\
T_{5.3} &=& [1, 1, ry_1^2-(1+r^2)y_1y_2+ry_2^2, 1, rx_1^2+(1+r)^2(x_1y_1+y_1^2), 1] \\
T_{5.4} &=& [1, 1, (y_1^2+y_2^2)^2, 1, (x_1^2+2x_1y_1+2y_1^2)^2, 1] \\
T_{5.5} &=& [1, 1, y_1^4+y_2^4+(2+\phi)(y_1^3y_2+y_2^3y_1)+(4+\phi)y_1^2y_2^2, 1, x_1^4+(2-\phi)(y_1^4 \\
&&\hspace{6.5cm} +x_1^3y_1+2y_1^3x_1+2y_1^2x_1^2), 1] \\
T_{5.6} 
&=& [1, x_1^2+x_1y_1+(2-\gamma)y_1^2, y_1^2+y_1y_2\gamma +y_2^2, 1, x_1^2+(2-\gamma)(x_1y_1+y_1^2), \\
&&\hspace{8cm}  x_1^2-x_1x_2\gamma +x_2^2],
\end{eqnarray*}
with $\gamma^3=3$ and $\phi(\phi-1)=1$.  Note that $H_{5.3}=H_5T_{3.3}$ and,\
both $H_{2.7}$ and $H_{4.9}$ divide $H_{5.6}$.

The equation
\begin{equation} \tag*{5.1}
\vec{u_t}{v_t}=\vec{eu^2+fuv+gv^2}{v_5+hv^2+iuv+ju^2}
\end{equation}
has approximate symmetries at orders $m\equiv 5,25 \mod 30$ with $c=0$.
The equation
\begin{equation} \tag*{5.2}
\vec{u_t}{v_t}=\vec{u_5+eu^2+fuv+gv^2}{v_5+hv^2+iuv+ju^2}
\end{equation}
has approximate symmetries at orders $m\equiv 1,5 \mod 6$ with $c=d=1$.
The equation
\begin{equation} \tag*{5.3}
\vec{u_t}{v_t}=\vec{\begin{split}\frac{1+r^5}{(1+r)^5}u_5+eu^2+f((r^4-r^3+r^2-r+1)(u_1v_3+2u_2v_2 \\
+2u_3v_1 +u_4v)-r(r^2+r+1)uv_4) +g((1+r^2)v_1^2 \\
+2(r^2+r+1)vv_2) \end{split}}{\begin{split} v_5+hv^2
+i((1+r)^2(uv_2+u_1v_1)+(r^2+r+1)u_2v) \\
+j(2r(r^2 +r+1)uu_4 +2(r^4+3r^3+5r^2+3r+1)u_1u_3 \\
+(r^4+5r^3 +7r^2+5r+1)u_2^2 )\end{split}}
\end{equation}
has approximate symmetries, with $c/d=(1+r^m)/(1+r)^m$, at orders $m \equiv 1,5 \mod 6$.
The equation
\begin{equation} \tag*{5.4}
\vec{u_t}{v_t}=\vec{-\frac{1}{4}u_5+eu^2+f(u_4v+2u_3v_1+2u_2v_2+u_1v_3+uv_4)
+gv^2}{v_5+hv^2+iuv+j(2uu_4+6u_1u_3+5u_2^2)}
\end{equation}
has approximate symmetries, with $c/d=(-1)^{(m-1)/4}2^{(1-m)/2}$, at orders $m \equiv 1,5 \mod 12$. 
The equation, with $\phi(\phi-1)=1$,
\begin{equation} \tag*{5.5}
\vec{u_t}{v_t}=\vec{\begin{split} -(4+5\phi) u_5 + eu^2
+ f((4-\phi)uv_4+11(u_1v_3+2u_2v_2 \\ +2u_3v_1+u_4v)) +gv^2 \end{split}}{
v_5 + hv^2+iuv+j(2\phi uu_4-2u_1u_3+(2\phi-1)u_2^2)}
\end{equation}
has approximate symmetries at orders $m\equiv 5, 25 \mod 30$, with
$c/d=-(1+F_{m-1}+ F_{m}\phi)$, where $F_m$ denotes the $m$-th Fibonacci number. 
And, the equation, with $\gamma^2=3$,
\begin{equation} \tag*{5.6}
\vec{u_t}{v_t}=\vec{\begin{split} (26 + 15 \gamma)u_5 + e u^2 + f(u_2v +u_1v_1+(\gamma-1)uv_2) \\
+g ( 4vv_2 + (3+\gamma)v_1^2)\end{split} }{\begin{split} v_5 + hv^2 + i(u_1v_1+uv_2-(1+\gamma)u_2v)\\
+ j(4uu_2+(3-\gamma)u_1^2 \end{split}}
\end{equation}
has approximate symmetries at orders $m\equiv 1,5 \mod 6$, with
$c/d$ given by equation (\ref{cd}).

\subsection{Higher order}
Define
\[
H_7=H_5[x_1^2+x_1x_2+x_2^2,1,1,y_1^2+y_1y_2+y_2^2,1,1],
\]
and, for convenience, $H_{2n}=H_2$, $H_{2n+1}=H_k$, where $n\equiv k$ mod $6$ and $k\in\{3,5,7\}$.
We first list the divisors that have similar structure at infinitely many higher orders. After that
we consider the exceptional cases.

\renewcommand{\theenumi}{\roman{enumi}}
\subsubsection*{The case $a=0$}
\begin{enumerate}
\item The tuple $\G_n[0,1]$ has the following divisor in $\H^6_n$
\begin{equation}
[0, y_1^n, y_1^n+y_2^n, X, y_1^n-(x_1+y_1)^n, (x_1+x_2)^n].
\end{equation}
where $X$ is the fourth component of $H_n$. It divides $\G_m[0,1]$ when
\[
m\equiv \left\{ \begin{array}{ll}
n \text{ mod } 2n, & n \text{ even, or } n\equiv 3 \text{ mod } 6, \\
n \text{ mod } 6n, & n\equiv 1 \text{ mod } 6, \\
n,5n \text{ mod } 6n, & n\equiv 5 \text{ mod } 6.
\end{array} \right.
\]
\end{enumerate}
\subsubsection*{The case $n$ is even}
Let $n>2$ be even. We have the following divisors in $\H_n^6$.
\begin{enumerate}
\setcounter{enumi}{1}
\item For any primitive ($n-1$)-st root of unity $r$
\begin{equation} \label{ev1}
H_2[1,1,(y_1-ry_2)^2(ry_1-y_2)^2,1,1,1]
\end{equation}
divides $\G_m[(1+r)^{1-m},1]$ when $m\equiv 1$ mod $n-1$.

\item Let one of $\mu,\nu$ be an $n$-th root of unity, and the other
a primitive $n$-th root of unity, such that $(\mu-1)(\nu-1)(\mu-\nu)(\mu\nu-1)\neq 0$.
Then, with $r=\nu(\mu-1)/(\nu-1)$
\[
H_2[1,1,(y_1-ry_2)(ry_1-y_2)(y_1-\bar{r}y_2)(\bar{r}y_1-y_2),1,1,1]
\]
divides $\G_m[1+r^m,(1+r)^m]$ when $m\equiv 0$ mod $n$.

\item For any primitive ($n-1$)-st root of unity $r$
\begin{equation} \label{ev2}
H_2[1,(y_1-(r-1)x_1)^2,1,1,1,1]
\end{equation}
divides $\G_m[(r-1)^{m-1},1]$ when $m\equiv 1$ mod $n-1$.

\item Let one of $\mu,\nu$ be a primitive $n$-th root of unity, and the other
a $n$-th root of unity, such that $(\mu-1)(\nu-1)(\mu-\nu)(\mu\nu-1)\neq 0$.
Then, with $r=(\mu-\nu)/(\nu-1)$
\[
H_2[1,(y_1-rx_1)(y_1-\bar{r}x_1),1,1,1,1]
\]
divides $\G_m[r^m,(1+r)^m-1]$ when $m\equiv 0$ mod $n$.

\item Let one of $\mu,\nu$ be a primitive $2n$-th root of unity, and the other
a primitive $2n$-th root of unity which is not a $n$-th root of unity, such that $(\mu-\nu)(\mu\nu-1)\neq 0$. Then, with $r=(\nu-\mu)/(\mu-1)$
\[
H_2[1,(y_1-rx_1)(y_1(1+\bar{r})+\bar{r}x_1),1,1,1,1]
\]
divides $\G_m[r^m,(1+r)^m-1]$ when $m\equiv n$ mod $2n$.

\item Let either $\mu$ be a primitive $n$-th root of unity and $\nu$ be a $2n$-th root
of unity which is not a $n$-th root of unity, or let $\mu\neq 1$ be a $n$-th root
of unity and $\nu$ a primitive $2n$-th root of unity. Then, with $r=-\nu(\mu-1)/(\nu-1)/\mu$
\[
H_2[1,1,(y_2-ry_1)(ry_2-y_1),1,1,\bar{r}x_1+(1+\bar{r})y_1 ,1]
\]
divides $\G_m[1+r^m,(1+r)^m]$ when $m\equiv n$ mod $2n$.
\end{enumerate}

\subsubsection*{The case $n$ is odd}
Let $n>3$ be odd. We have the following divisors in $\H_n^6$.
\begin{enumerate}
\setcounter{enumi}{7}
\item For any primitive ($n-1$)-st root of unity $r$ define
\[
Q_n(r)=[1,1,(y_1-ry_2)^2(ry_1-y_2)^2,1,(x_1+(1+r)y_1)^2(rx_1+(1+r)y_1)^2,1].
\]
The tuple $H_nQ_n(r)$ divides $\G_m[(1+r)^{1-m},1]$ when
\[
m\equiv \left\{ \begin{array}{ll}
1 \text{ mod } n-1, & n \equiv 1,3 \text{ mod } 6, \\
1,n \text{ mod } 3(n-1), & n\equiv 5 \text{ mod } 6.
\end{array} \right.
\]

\item Let one of $\mu,\nu$ be a $n$-th root of unity, and the other
a primitive $n$-th root of unity, such that $(\mu-1)(\nu-1)(\mu-\nu)(\mu\nu-1)\neq 0$.
Then, with $r=\nu(\mu-1)/(\nu-1)$ we define
\[
\begin{split}W_n(r)=[1,1,(y_1-ry_2)(ry_1-y_2)(y_1-\bar{r}y_2)(\bar{r}y_1-y_2),1,(x_1+(1+\bar{r})y_1) \\
\hspace{17mm} (\bar{r}x_1+(1+\bar{r})y_1)(x_1+(1+\bar{r})y_1)(\bar{r}x_1+(1+\bar{r})y_1),1].
\end{split}
\]
The tuple $H_nW_n(r)$ divides $\G_m[1+r^m,(1+r)^m]$ when
\[
m\equiv \left\{ \begin{array}{ll}
n \text{ mod } 6n, & n \equiv 1 \text{ mod } 6, \\
n \text{ mod } 2n, & n\equiv 3 \text{ mod } 6, \\
n,5n \text{ mod } 6n, & n\equiv 5 \text{ mod } 6.
\end{array} \right.
\]
\end{enumerate}

We also get new highest degree divisors in $\H_n^6$, with $n$ odd, from primitive
$p$-th roots of unity with $p<n$. This happens when $n\equiv 1$ mod $p$ (or
$n\equiv 0$ mod $p$) and $n\equiv k$ mod $6$ with $k\in\{3,5,7\}$, such that there is no
$1<q<n$ with $q\equiv 1$ mod $p$ (or $q\equiv 0$ mod $p$), and $q\equiv l$ mod $6$ with $l\in\{3,5,7\}$
and $l\geq k$. For example, the tuple $T_{5.4}$ divides a $\G_m$ at $m \equiv 1$ mod $4$. The tuple $H_{5.4}=H_5T_{5.4}$
divides a $\G_m$ at $m\equiv 1,5$ mod $12$, see equation 5.4. We have $13\equiv 1$ mod $4$ and $13\equiv 7$ mod
$6$. There is no $1<q<13$ such that $q\equiv 1$ mod $4$ and $q\equiv l$ mod $6$, with $l\in{3,5,7}$
and $l\geq7$. Thus, the tuple $H_7T_{5.4}\in\H_{13}$ divides a $\G_m$ at $m\equiv 1$ mod $12$. We have $9\equiv 1$ mod $4$
and $9\equiv 3$ mod $6$, but there is $q=5$ such that $q\equiv 1$ mod $4$ and $q\equiv 5$ mod $6$, and $5\geq3$.
Therefore the tuple $H_3T_{5.4}$ is not in $\H_9$. It is in $\H_5$ but it does not have maximal degree.

Also, we may have $p$ odd. If the tuple (\ref{ev1}) divides $\G_m[(1+r)^{1-m},1]$
when $m\equiv 1$ mod $p$, then, according to Proposition \ref{1-121}, $Q^{p+1}_{(r)}$ divides
$\G_{2p+1}[(1+r)^{2p},1]$. This always give us a new highest degree tuple in $\H_{2p+1}$.
On the other hand, from case iii one can conclude, using Propositions \ref{101} and \ref{1-121}
and inverting $a\rightarrow 1/a$, that $Q_{p+1}(-r)$ divides $\G_{2p+1}[(1+r)^{2p},1]$.
This we knew already, since when $r$ is a primitive $p$-th root of unity, with $p$ odd, then
$-r$ is a $2p$-th root of unity, cf. case viii.

In general one can show the following.
\begin{itemize}
\item Suppose $n \equiv 0$ mod $p$ and $n\equiv k$ mod $6$ with $k\in\{3,5,7\}$,
such that there is no $q<n$ with $q\equiv 0$ mod $p$, and $q\equiv l$ mod $6$ with
$l\in\{3,5,7\}$ and $l\geq k$. Then $p$ is odd. And,
\begin{equation} \label{c1}
n=\left\{ \begin{array}{ll}
p, & p\equiv 1,3 \text{ mod } 6, \\
p,5p & p\equiv 5 \text{ mod } 6.
\end{array} \right.
\end{equation}
\item Suppose $n \equiv 1$ mod $p$ and $n\equiv k$ mod $6$ with $k\in\{3,5,7\}$,
such that there is no $q<n$ with $q\equiv 1$ mod $p$, and $q\equiv l$ mod $6$ with
$l\in\{3,5,7\}$ and $l\geq k$. Then
\begin{equation} \label{c2}
n=\left\{ \begin{array}{ll}
p+1 & p\equiv 0 \text{ mod } 6, \\
p+1,2p+1,4p+1,6p+1 & p\equiv 1 \text{ mod } 6, \\
p+1,2p+1,3p+1 & p\equiv 2 \text{ mod } 6, \\
p+1,2p+1 & p\equiv 3 \text{ mod } 6, \\
p+1,3p+1 & p\equiv 4 \text{ mod } 6, \\
p+1,2p+1,6p+1 & p\equiv 5 \text{ mod } 6.
\end{array} \right.
\end{equation}
\end{itemize}
We can now describe all highest degree divisors at odd order $n$, involving the tuples $Q,W$. For odd $n$ there are
the cases viii and ix described above. They corresponds to the cases
$n=p+1$ in (\ref{c2}), and $n=p$ in (\ref{c1}), respectively. Furthermore we have:
\begin{enumerate}
\setcounter{enumi}{9}
\item If $n=5p$ and $p\equiv 5$ mod $6$, then $H_nW_p(r)$ divides $\G_m[1+r^m,(1+r)^m]$ when $m\equiv 5p$ mod $6p$.
\item If $n\equiv 1,3$ mod $6$ and equation (\ref{c2}) holds for certain $p$, then $H_nQ_p(r)$ divides
$\G_m[1+r^m,(1+r)^m]$ when $m\equiv 1$ mod $n-1$.
\item Let $n\equiv 5$ mod $6$. If $n=2p+1$, $p\equiv 5$ mod $6$ then $H_nQ_p(r)$ divides
$\G_m[1+r^m,(1+r)^m]$ when $m\equiv 1,n$ mod $3(n-1)$. And if $n=2p+1$, $p\equiv 2$ mod $6$, or
$n=4p+1$, $p\equiv 1$ mod $6$, then $H_nQ_p(r)$ divides $\G_m[1+r^m,(1+r)^m]$ when $m\equiv 1,n$ mod $3(n-1)/2$.
\end{enumerate}

\subsubsection*{Exceptional highest degree divisors, $n\geq 7$}
\begin{enumerate}
\setcounter{enumi}{12}
\item The tuple $\G_7[1,1]\in\H_7$ divides $\G_m[c,d]$ with $m\equiv 1$ mod $6$ and $c=d=1$.
\item The tuple $H_7T_{3.3}\in\H_7$ divides $\G_m[1+r^m,(1+r)^m]$ with $m\equiv 1$ mod $6$.
\item The tuple $H_7T_{3.4}\in\H_7$ divides $\G_m[c,d]$ with $m\equiv 1,7,13,19$ mod $30$ and
$c,d$ given by equation \ref{fib}.
\item The tuple $H_7T_{5.6}\in\H_7$ divides $\G_m[c,d]$ with $m\equiv 1$ mod $6$ and
$c,d$ given by equation \ref{cd}.
\item Define
\begin{align*}
Z=[1, (x_1^2+x_1y_1+(1+\phi)y_1^2)^2, (y_1^2+(1-\phi)y_1y_2+y_2^2)^2, 1, \ &\\
(x_1^2+(1+\phi)(x_1y_1+y_1^2))^2, (x_1^2+\phi x_1x_2+x_2^2)^2].&
\end{align*}
Then $H_5Z\in\H_{11}$ divides $\G_m[c,d]$ with $m\equiv 1,11$ mod $30$ and $c/d$ given by
equation (\ref{fib}). And $H_7Z\in \H_{31}$ divides $\G_m[c,d]$ with $m\equiv 1$ mod $30$.

\item The tuple
\begin{align*}
H_7 [1, (x_1^2+x_1y_1+(2-\gamma)y_1^2)^2, (y_1^2+\gamma y_1y_2+y_2^2)^2, 1, \ &\\
(x_1^2+(2-\gamma)(x_1y_1+y_1^2))^2, (x_1^2-\gamma x_1x_2+x_2^2)^2]\in \H_{13}&
\end{align*}
divides $\G_m[c,d]$ with $m\equiv 1$ mod $12$ and $c/d$ given by
equation (\ref{cd}).
\end{enumerate}

\section{Concluding remarks} \label{cr}
In the formal symmetry approach \cite{MSS}, as well as in the
computer-assisted schemes \cite{Fou1, TsWo}, not knowing the ratios of eigenvalues strongly complicates
the classification of integrable equations. There the ratios are obtained, if possible at all,
at the very last stage of the calculations. We hope that the a priori knowledge provided in this
article will be an impetus to complete the classification.

Usually, in classification programs one considers homogeneous equations. A 2-component equation
$(u_t,v_t)=K$ is {\em homogeneous of weighting $\lambda$} if $K$ is an eigenvector of $\L(\sigma_x+\lambda_1\sigma_u
+\lambda_2\sigma_v)$, where $\sigma_x=(xu_1,xv_1)$ counts the number of derivatives. We have compactly provided
a list of nonhomogeneous equations. A complete list of homogeneous equations can be obtained from our list by
multiplying the symbolic quadratic parts with appropriate tuples of polynomials. And Lemma \ref{lem2} can be
used to determine all symmetries of those equations.

A classification of second order integrable 2-component evolution equations has been given in \cite{SW01a}.
The lemmas 6.3, 6.4, 6.5, and 6.6, proved in there, are special cases ($n=2$) of Proposition \ref{1-121},
\ref{1-122}, \ref{1021}, and \ref{1-110}, respectively. In \cite{SW01a} the full analysis of higher order
symmetry conditions was carried out completely. From this it follows that all second order integrable
equations with quadratic parts are derived from equations 2.3, 2.4, and 2.5. Note that the authors
of \cite{SW01a} excluded equations of type 2.1 and 2.2.

A classification of third order 2-component evolution equations with weighting $(2,2)$ and symmetries
of order 5, 7, or 9, is given in \cite{Fou1}. Of the five equations listed \cite[Theorem 3.3]{Fou1}, two
are nonlinear injective and have diagonalizable linear part:
\begin{equation} \label{-2}
\vec{u_t}{v_t}=\vec{u_3+uu_1+vv_1}{-2v_3-uv_1}
\end{equation}
and
\begin{equation} \label{43}
\vec{u_t}{v_t}=\vec{4u_3 + 3v_3 + 4uu_1+vu_1+2uv_1}{3u_3 + v_3 - 2vv_1-4vu_1-2uv_1}.
\end{equation}
When put in Jordan form the ratio of the coefficients of the linear part of equation (\ref{43}) becomes $a/b=-3\phi-2$,
where $\phi$ is the golden ratio. Our diophantine approach perfectly explains the 'unusual' symmetry pattern. We
remark that the conjugate of $\phi$ gives rise to another equation with $a/b=-3(1-\phi)-2=1/(-3\phi-2)$, which can
be obtained by interchanging $u$ and $v$. A similar remark holds for all equations derived from the 'symmetric'
Propositions \ref{1-122} and \ref{1021}. For example, by interchanging $u$ and $v$ in equation 4.11 we get an equation
with $a/b=1/(7+4\gamma)=7-4\gamma$.

According to \cite[Theorem 3.3]{Fou1} we have the following. {\em A non-decouplable
fifth order two component equation in the KDV weighting, possessing a generalized symmetry of order 7 can
be reduced by a linear change of variables to a symmetry of lower order equations, or to the Zhou-Jiang-Jiang
equation,}
\[
\vec{u_t}{v_t}=\vec{\begin{split} u_5-5(2uu_3+5u_1u_2)+15(2vv_3+3v_1v_2)+20u_1u^2 \\
-30(u_1v^2+2uvv_1)\end{split}}{\begin{split}-9v_5+5(2u_3v+7u_2v_1+9u_1v_2+6uv_3)-10(2uu_1v\\+2u^2v_1+3v^2v_1)\end{split}}.
\]
However, since the ratio of coefficients of the linear part $a/b=-1/9$ does not appear in our list, this equation
has an approximate symmetry at lower order. In fact, the equation has a genuine symmetry at third order. The
Zhou-Jiang-Jiang equation is in the hierarchy of the Drinfel'd-Sokolov type \cite{DS} equation
\begin{equation} \label{zjjh}
\vec{u_t}{v_t}=\vec{-3vv_1}{v_3-u_1v-2uv_1},
\end{equation}
which is linearly equivalent to a third order equation that appears in the same paper \cite[Equation (17)]{Fou1},
cf. \cite[Sections 3.2.1, 4.2.6]{TsWo}. The special value of the ratio $a/b=-1/2$ in equation (\ref{-2}) also does
not appear in our list and is due to higher grading constraints. At the end of \cite{MNW2} two fifth order systems
are given with ratios $(9-5\sqrt{3})(9+5\sqrt{3})^{-1}=26-15\sqrt{3}$ and $-1/9$. The first system derives from equation 5.6
and the latter from equation 5.3, with $r$ a primitive $6$-th root of unity.

We conclude with a more philosophical remark and some ideas on future research. The concept of generalized symmetry
really is about local symmetry. The (inverse) Gel'fand-Diki{\u\i} transformation translates polynomials in the symbols
$x,y$ into local differential functions,
that is expressions in $u,v$ and their derivatives. A question arises: can we also translate rational functions in the
symbols $x,y$? The answer is yes. One could think of non-local variables $u_i,v_i$ with $0 > i \in\Z$. Here a negative
index indicates integration and $D_x$ would be such that $D_x(u_{i-1})=u_i$ for all $i$. We can expand rational
functions in Taylor-series, which are transformed into non-local differential sums. For example, consider the
rational function $\widehat{F}=1/(x_1+x_2)$. Its symmetric Taylor-series is
\[
\widehat{F}=\frac{1}{2} \sum_{k=0}^\infty (-1)^k (\frac{x_1^k}{x_2^{k+1}} + \frac{x_2^k}{x_1^{k+1}})
\]
which is transformed into the non-local object
\[
F=\sum_{k=0}^\infty (-1)^k u_ku_{-k-1},
\]
and we have $D_xF=u^2$. In this non-local setting every equation has a symmetry at any order. For example,
the equation, with $a\neq 1$,
\[
\vec{u_t}{v_t}=\vec{au_1+(1-a)uv}{v_1+(1-a)uv},
\]
and its approximate symmetries (but the ones in $\AR^{2,0}\otimes\AR^{0,2}$), are in the approximate hierarchy of the
zeroth order non-local equation
\[
\vec{u_t}{v_t}=\vec{u + uv_{-1}}{v - u_{-1}v}.
\]
Still, there would be a quest for equations, or symmetries, that are in certain sense close to local.

Since the pioneering work of Sanders and Wang \cite{MR99g:35058}, apart from extending the result to equations
with more more components \cite{MR1829636,SW01a,PH02}, the symbolic method has been further developed
in order to classify non-commutative \cite{OW}, non-evolutionary \cite{psa,MNW,NW}, non-local \cite{psa,MN},
and multi-dimensional equations \cite{JPW}. In classifying non-local equations the concept of quasi-locality,
introduced in \cite{MY} is the key idea. Only certain types of non-localities are allowed by considering
different extensions of the ring of differential polynomials. `Nnon-evolutionary' equations are treated as
evolution equations with possible non-localities. For example, a Boussinesq type scalar equation
\[
u_{tt}=K(u,u_x,u_{xx}\ldots,u_t,u_{xt},\ldots
\]
can be represented as a two-component evolution equation,
\[
\vec{u_t}{v_t} = \vec{v}{K(u,u_x,\ldots,v,v_x,\ldots)}.
\]
However, different evolutionary representation may exist, and some might have local symmetries, whereas others
might posses non-local symmetries \cite{MNW}. Another example is the Camassa-Holm type equation
\[
\vec{m_t}{u_{xx}} = \vec{c m u_x + um_x}{u-m},
\]
which is integrable when $c=2$ \cite{CH} and $c=3$ \cite{DP}. By eliminating the variable $u$ this equation is
written as
\[
m_t=c m \Delta m_x + m_x \Delta m
\]
where $\Delta=(1-D_x^2)^{-1}$ is a non-local operator. Its symmetries are quasi-local expressions
in $D_x$-, and $\Delta$-derivatives of $m$ \cite{psa,MNW2}. 

Another interesting problem would be to classify non-evolutionary equations as they are, that is, to apply the
symbolic method for polynomials in both $x$ and $t$ derivatives, developed in \cite{JPW}. In the setting of
non-evolutionary equations, there is a clear distinction between the equation and its symmetries.
When $\Delta=0$ represents the equation, a function
$Q$ is an infinitesimal symmetry if the prolongation of $Q$ acting on $\Delta$ vanishes modulo $\Delta$
\cite[Theorem 2.31]{Olv}. For example, we have $u_{txx}-u_tu_x$ as a symmetry
of the Boussinesq equation $u_{tt}=u_{xxxx}-2u_xu_{xx}$, but not vise versa.
Recently we formulated an equivalent symmetry condition in terms of Lie-brackets, in the
setting of passive orthonomic systems \cite{SL}. This could play a role in the general classification problem.
We also found that certain hierarchies of evolution equations appear as symmetries of non-evolutionary equations.
For example, it seems that all symmetries of the Sawada-Kotera equation \cite{SK}
\[
u_t=u_5-5u_3u_1+\frac{5}{3}w_1^3
\]
are symmetries of $u_{txx}=u_tu_x$, which is a special case of Ito's equation \cite{PV}. Based on Lax-pairs
a similar observation was made in \cite{HW}.

\subsection*{Acknowledgment} This research has been funded by the Australian Research Council
through the Centre of Excellence for Mathematics and Statistics of Complex Systems. The work is based
on research done at the Faculty of Sciences, Department of Mathematics, {\it vrije} Universiteit \ {\it amsterdam}.
I thank Frits Beukers, Jan Sanders and Jing Ping Wang for the research has been built on their ideas and
encouragement. I thank Jaap Top for his interest, Bob Rink for some nice discussions, one anonymous referee for
clarifying the full force of Theorem \ref{smlt}, and another for pointing out some incompleteness.

\end{document}